\documentclass[12pt]{extarticle}
\usepackage{amsmath}

\usepackage{amsfonts}

\usepackage{graphicx,subfigure}
\usepackage{amsmath,amssymb,amsthm,amsfonts,graphicx,shorttoc,textpos,caption,here,titlesec,stmaryrd,bbold, amssymb,mathrsfs,eqnarray,upgreek}
\usepackage{verbatim,enumerate,hyperref,dsfont,fancyhdr,setspace,array,multirow,supertabular,multicol,subfigure,enumitem}
\usepackage{algorithm}
\usepackage{algorithmic}

\input{tcilatex}

\begin{document}

\title{A sequential design for extreme quantiles estimation under binary
sampling.}
\author{Michel Broniatowski and Emilie Miranda \\
LPSM, CNRS UMR 8001, Sorbonne Universite, Paris}
\maketitle

\begin{abstract}
We propose a sequential design method aiming at the estimation of an extreme quantile based on a sample of dichotomic data 
corresponding to peaks over a given threshold. This study is motivated by an industrial challenge in material reliability and consists
in estimating a failure quantile from trials whose outcomes are reduced to indicators of whether the specimen have failed at the
 tested stress levels. The solution proposed is a sequential design making use of a splitting approach, decomposing the target
probability level into a product of probabilities of conditional events of higher order. The method consists in gradually targeting 
the tail of the distribution and sampling under truncated distributions. The model is GEV or Weibull, and sequential estimation of its parameters
 involves an improved maximum likelihood procedure for binary data, due to the large uncertainty associated with such a restricted information.
\end{abstract}

Consider a non negative random variable $X$ with distribution function $G$.$%
\ $\ Let $X_{1},..,X_{n}$ be $n$ independent copies of $X.$ The aim of this
paper is to estimate $q_{1-\alpha }$, the $\left( 1-\alpha \right) $%
-quantile of $G$ when $\alpha $ is much smaller than $1/n.$ We therefore aim
at the estimation of so-called extreme quantiles. This question has been
handled by various authors, and we will review their results somehow later.
The approach which we develop is quite different since we do not assume that
the $X_{i}$'s \ can be observed. For any threshold $x$, we define the r.v. 
\[
Y=\left\{ 
\begin{array}{l}
1\text{ if }X\leq x \\ 
0\text{ if }X>x%
\end{array}%
\right. 
\]%
which therefore has a Bernoulli distribution with parameter $G(x).$ We may
choose $x$, however we do not observe $X$, but merely $Y.$ Therefore any inference on 
$G$ suffers from a severe loss of information. This kind of setting is common in industrial statistics: When exploring the strength of a material, or of a
bundle, we may set a constraint $x$, and observe whether the bundle
breaks or not when subjected at this level of constraint. 

In the following, we will denote $R$ the resistance
of this material, we observe $Y.$ Inference on $G$ can be performed for large $%
n$ making use of many thresholds $x.$ Unfortunately\ such a procedure will
not be of any help for extreme quantiles. To address this issue, we will consider a design of experiment enabling to progressively
characterize the tail of the distribution by sampling at each step in a more extreme region of the density. It will thus be assumed in the following that we are able to observe $Y$ not only when $R$ follows $G$ but also when $R$ follows
the conditional distribution of $R$ given $\{ R>x\}.$ In such a case we will be
able to estimate $q_{1-\alpha }$ even when $\alpha <1/n$ where $n$
designates the total number of trials.\ In material sciences, this amounts to
consider trials based on artificially modified materials; in the case when
we aim at estimation of extreme upper quantiles, this amounts to strengthen
the material. We would consider a family of increasing thresholds $%
x_{1},..,x_{m}$ and for each of them realize $K_{1},..,K_{m}$ trials, each
block of iid realizations $Y$'s being therefore functions of the
corresponding unobserved $R$'s with distribution $G$ conditioned upon $%
\{R>x_{l}\}$, $1\leq l\leq m.$ 
design which allows for the estimation of extreme quantiles.

The present setting is therefore quite different from that usually
considered for similar problems under complete information. As sketched
above it is specifically suited for industrial statistics and reliability
studies in the science of materials.\ 

From a strictly statistical standpoint, the above description may also be
considered when the distribution $G$ is of some special form, namely when
the conditional distribution of $R$ given $\{R>x\}$ has a functional form which
differs from that of $G$ only through some changes of the parameters. In
this case, simulation under these conditional distributions can be performed
for adaptive choice of the thresholds $x_{l}$'s, substituting the above
sequence of trials. This sequential procedure allows to estimate iteratively
the initial parameters of $G$ and to obtain $q_{1-\alpha }$ combining
corresponding quantiles of the conditional distributions above thresholds, a
method named splitting.\ In this method, we will choose sequentially the $%
x_{l}$'s in a way that $q_{1-\alpha }$ will be obtained easily from the last
distribution of $x$ conditioned upon $\{R>x_{m}\}.$

In safety issues or in pharmaceutical control, the focus is usually set on the
behavior of a variable of interest (strength, maximum tolerated dose) for
small (or even\ very small) levels. In these settings the above
considerations turn to be equivalently stated through a clear change of
variable, considering the inverse of the variable of interest. As an example
which is indeed at the core of the motivation for this paper, and in order
to make this approach more intuitive, we first sketch briefly the industrial
situation which motivated this work in Section  \ref{IndContext}. We look at a
safety property, namely thresholds $x$ which specify very rare events, typically failures under very
small solicitation.

As stated above, the problem at hand is the estimation of very small
quantiles. Classical techniques in risk theory pertain to large quantiles
estimation. For example, the Generalized Pareto Distribution, to be referred
to later on, is a basic tool in modeling extreme risks and exceedances over
thresholds. Denoting $R$ the variable of interest and $\widetilde{R}:=1/R,$
then obviously, for $x>0$, $\left\{ R<x\right\} $ is equivalent to $\left\{
\widetilde{R}>u\right\} $ with $u=1/x$. In this paper we will therefore make
use of this simple duality, stating formulas for $R,$ starting with
classical results pertaining to $\widetilde{R}$ when necessary. Note that
when $q_{\alpha }$ designates the $\alpha -$quantile of $R$ and respectively 
$\widetilde{q}_{1-\alpha }$ the $\left( 1-\alpha \right)-$quantile of $%
\widetilde{R}$, it holds $q_{\alpha }=1/\widetilde{q}_{1-\alpha }.$The
resulting notation may seem a bit cumbersome; however the reader accustomed to
industrial statistics will find it more familiar. 

This article is organized as follows. Section \ref{IndContext} formalizes the problem
in the framework of an industrial application to aircraft industry. In Section \ref{revLit}, a short survey of extreme quantiles estimation and of existing designs of experiment are studied as well as their applicability to extreme quantiles estimation. Then, a new procedure is proposed in Section \ref{Splitting} and elaborated for a Generalized Pareto model. An estimation procedure is detailed and evaluated in Section \ref{EstimationProc}. Then an alternative Weibull model for the design proposed  is presented in Section \ref{WeibullModel}. Lastly, Sections \ref{model_selection_missp} and \ref{Perspectives} provide a few ideas discussing model selection and behavior under misspecification as well as hints about extensions of the models studied beforehand.

\section{Industrial challenge}\label{IndContext} 

\subsection{Estimation of minimal allowable stress in material fatigue}
In aircraft industry, one major challenge is the characterization of extreme
behaviors of materials used to design engine pieces. Especially, we will
consider extreme risks associated with fatigue wear, which is a very
classical type of damage suffered by engines during flights. It consists in
the progressive weakening of a material due to the application of cyclic loadings a large number of times
that can lead to its failure. As shown in Figure \ref%
{cyclefatigue}, a loading cycle is defined by several quantities: the
minimal and maximal stresses $\sigma _{\min }$ et $\sigma _{\max }$, the
stress amplitude $\sigma _{a}=\frac{\sigma _{\max }-\sigma _{\min }}{2}$,
and other indicators such as the stress ratio $\frac{\sigma _{\min }}{\sigma
_{\max }}$. 

\begin{center}
\includegraphics[scale=0.05]{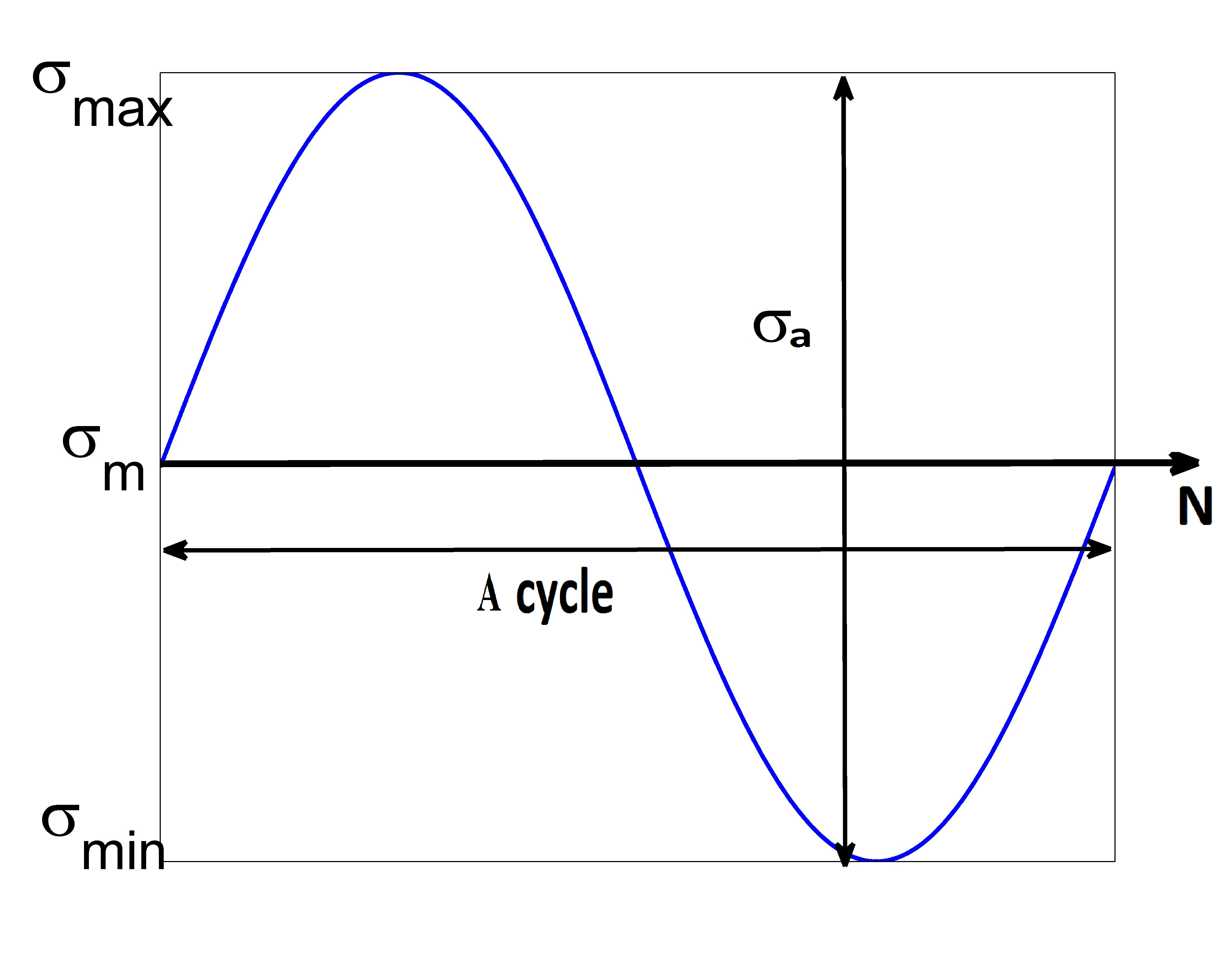} 
\captionof{figure}{Loading cycle on a material}\label{cyclefatigue}
\end{center}

\bigskip 

The fatigue strength of a given material is studied through experimental
campaigns designed at fixed environmental covariates to reproduce flight
conditions. The trials consist in loading at a given stress level a
dimensioned sample of material up to its failure or the date of end of
trial. The lifetime of a specimen is measured in terms of number of cycles
to failure, usually subject to right censoring.

\begin{figure}[h]
\begin{center}
\includegraphics[scale=0.9]{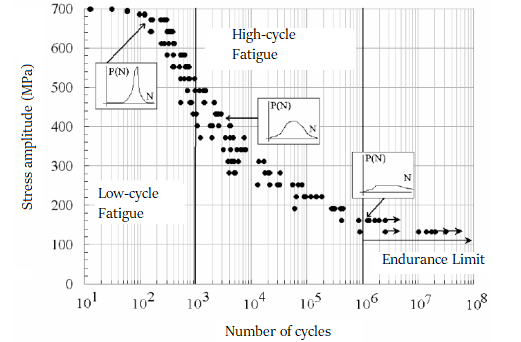}
\end{center}
\caption{S-N curve}
\label{wohler}
\end{figure}
\bigskip

The campaign results are then used to study fatigue resistance and are
represented graphically in an S-N scale (see figure \ref{wohler}). S-N curves
highlight the existence of three fatigue regimes. Firstly, low cycle fatigue
corresponds to short lives associated with high levels of stress. Secondly,
during high cycle fatigue, the number of cycles to failure decreases
log-linearily with respect to the loading. The last regime is the endurance
limit, in which failure occurs at a very high number of cycles or doesn't
occur at all. We will focus in the following on the endurance limit, which
is also the hardest regime to characterize since there is usually only few
and scattered observations.

In this framework, we are focusing on minimal risk. The critical
quantities that are used to characterize minimal risk linked to fatigue
damage are failure quantiles, called in this framework allowable stresses at
a given number of cycles and for a fixed level of probability. Those
quantiles are of great importance since they intervene in decisions pertaining 
engine parts dimensioning, pricing decisions as well as maintenance
policies.

\subsection{Formalization of the industrial problem}

The aim of this study is to propose a new design method for the
characterization of allowable stress in very high cycle fatigue, for a very
low risk $\alpha $ of order $10^{-3}$. We are willing to obtain a precise
estimation method of the $\alpha -$failure quantile based on a minimal
number of trials.

Denote $N$ the lifetime of a material in terms of number of cycles to
failure and $S$ the stress amplitude of the loading, in MPa. Let $n_{0}$ be
the targeted time span of order $10^{6}-10^{7}$ cycles.

Define the allowable stress $s_{\alpha }$ at $n_{0}$ cycles and level of
probability $\alpha $ $=10^{-3}$ the level of stress that guarantee that the
risk of failure before $n_{0}$ does not exceed $\alpha $: 
\begin{equation}
s_{\alpha }=\sup \left\{ s:\mathbb{P}(N\leq n_{0}|S=s)\leq \alpha \right\} 
\label{pb1}
\end{equation}

\bigskip 


\bigskip
We will now introduce a positive r.v. $R=R_{n_{0}}$ modeling the resistance
of the material at  $n_{0}$ cycles and homogeneous to the stress. $R$
is the variable of interest in this study and its distribution $\mathbb{P}$ is defined as: 
\begin{equation}
\mathbb{P}(R\leq s) = \mathbb{P}(N\leq n_{0}|S=s).  \label{loiR}
\end{equation}

Thus, the allowable stress can be rewritten as the $\alpha-$quantile of
the distribution of $R$,

\QTP{Body Math}
\begin{equation}
s_{\alpha }=q_{\alpha }=\sup \left\{ s:\mathbb{P}(R\leq s)\leq \alpha \right\}. 
\end{equation}

However, $R$ is not directly observed. Indeed, the usable data collected at
the end of a test campaign consists in couples of censored fatigue life -
stress levels $\left(\min (N,n_{0}),s\right)$ where $s$ is part of the design of the
experiment. The relevant information that can be drawn from those
observations to characterize $R$ is restricted to indicators of whether or
not the specimen tested has failed at $s$ before $n_{0}$. Therefore, the
relevant observations corresponding to a campaign of $n$ trials are formed
by a sample of variables $Y_{1},...,Y_{n}$ with for $1\leq i\leq n,$ 
\[
Y_{i}=\left\{ 
\begin{array}{l}
1\text{ if }R_{i}\leq s_{i} \\ 
0\text{ if }R_{i}>s_{i}%
\end{array}%
\right. 
\]

where $s_{i}$ is the stress applied on specimen $i.$\bigskip 


Note that the number of observations is constrained by industrial and
financial considerations; Thus $\alpha $ is way lower than $1/n$ and we are
considering a quantile lying outside the sample range.

\bigskip 

While we motivate this paper with the above industrial application, note
that this kind of problem is of interest in other domains, such as broader
reliability issues or medical trials through the estimation of the maximum tolerated
dose of a given drug.

\section{Extreme quantile estimation, a short survey}\label{revLit}

As seen above estimating the minimal admissible constraint raises two
issues; on one hand the estimation of an extreme quantile, and on the other
hand the need to proceed to inference based on exceedances under
thresholds.\ \ We present a short exposition of these two areas, keeping in
mind that the literature on extreme quantile estimation deals with complete
data, or data under right censoring.

\subsection{Extreme quantiles estimation methods}

%
%
Extreme quantile estimation in the univariate setting is widely covered in the literature when the variable of interest $X$ is either completely or 
partially observed.

The usual framework is the study of the $(1-\alpha)-$quantile of a r.v $X$, with very small $\alpha$.

The most classical case corresponds to the setting where ${x}_{1-\alpha}$ is drawn from a $n$ sample of observations $X_1,\dots X_n$. 
We can distinguish estimation of high quantile, where $x_{1-\alpha}$ lies inside the sample range, see Weissman 1978 \cite{Weissman} and Dekkers and al. 1989 \cite{dekkers1989},
and the estimation of an extreme quantile outside the boundary of the sample, see for instance De Haan and Rootz\'en 1993 \cite{deHann1993}. 
It is assumed that $X$ belongs to the domain of attraction of an extreme value distribution. The tail index of the latter is then estimated through maximum likelihood (Weissman 1978 \cite{Weissman})
or through an extension of Hill's estimator (see the moment estimator by  Dekkers and al. 1989 \cite{dekkers1989}).
Lastly, the estimator of the quantile is deduced from the inverse function of the distribution of the $k$ largest observations. 
Note that all the above references assume that the distribution has a Pareto tail. An alternative modeling has been proposed by  De Valk 2016 \cite{valk}
and De Valk and Cai 2018 \cite{valk2}, and consists in assuming a Weibull type tail, which enables to release some second order hypotheses on the tail.
This last work deals with the estimation of extreme quantile lying way outside the sample range and will be used as a benchmark method in the following sections.

Recent studies have also tackled the issue of censoring. For instance, Beirlant and al. 2007 \cite{beirlant2007} and Einmahl and al. 2008 \cite{Einmahl2008} proposed
a generalization of the peak-over-threshold method when the data are subjected to random right censoring and an estimator for extreme quantiles. The idea is to consider a consistent 
estimator of the tail index on the censored data and divide it by the proportion of censored observations in the tail. 
Worms and Worms 2014 \cite{worms2014} studied estimators of the extremal index based on Kaplan Meier integration and censored regression.

However the literature does not cover the case of complete truncation, i.e when only exceedances over given thresholds are observed. Indeed, all of the above 
are based on estimations of the tail index over weighed sums of the higher order statistics of the sample, which are not available in the problem of interest in this study.
Classical estimation methods of extreme quantiles are thus not suited to the present issue. 

In the following, we study designs of experiment at use in industrial contexts and their possible application to extreme quantiles estimation.
\subsection{Sequential design based on dichotomous data}

In this section we review two standard methods in the industry and in
biostatistics, which are the closest to our purpose.\ Up to our knowledge,
no technique specifically addresses inference for extreme quantiles.

We address the estimation of small quantiles, hence the events of interest
are of the form $\left( R<s\right) $ and the quantile is $q_{\alpha }$ for
small $\alpha .$

The first method is the \textit{staircase}, which is the present tool
used to characterize a material fatigue strength\textit{.}

\smallskip The second one is the \textit{Continual Reassessment Method (CRM)}
which is adapted for assessing the admissible toxicity level of a drug in
Phase 1 clinical trials.

Both methods rely on a parametric model for the distribution of the strength
variable $R.$ We have considered two specifications, which allow for simple
comparisons of performance, and do not aim at an accurate modelling in
safety.

\subsubsection{The Staircase method}

Denote $\mathbb{P}(R\leq s)=\phi (s,\theta _{0})$. Invented by Dixon and
Mood (1948 \cite{dixon}), this technique aims at the estimation of the 
parameter $\theta _{0}$ through sequential search based on data of
exceedances under thresholds. The procedure is as follows.

\textbf{Procedure}

Fix 
\begin{itemize}
\item[$\bullet $] The initial value for the constraint, $S_{ini}$, 
\item[$\bullet $]  The step $\delta >0$, 
\item[$\bullet $] The number of cycles $n_0$ to perform before concluding a trial,
\item[$\bullet $] The total number of items to be tested, $K$.
\end{itemize}
The first item is tested at level $s_{(1)}=S_{ini}$. The next item is tested at level $s_{(2)}=S_{ini}-\delta$ in case of failure and 
$s_{(2)}=S_{ini}+\delta$ otherwise. Proceed sequentially on the $K-2$ remaining specimen at a level increased (respectively decreased) by $\delta$ in case of survival (resp. failure).  The process is illustrated in figure \ref{staircase}.

Note that the proper conduct of the Staircase method relies on strong assumptions on the choice of the design parameters. Firstly, $S_{ini}$ has to be sufficiently close to the expectation of $R$ and secondly, $\delta $ has to lay between $0.5\sigma $ and $2\sigma $, where $\sigma$ designates the standard deviation of the distribution of $R$.

Denote $\mathbb{P}(R\leq s)=\phi (s,\theta _{0})$ and $Y_{i}$ the variable associated to the issue of the trial $i$, $1\leq
i\leq K$, where $Y_{i}$ takes value $1$ under failure and $0$ under no
failure, $Y_{i}=\mathds{1}_{N_{a}\leq n_{0}}\sim \mathcal{B}(\phi
(s_{i},\theta _{0}))$.

\begin{center}
\includegraphics[scale=0.7]{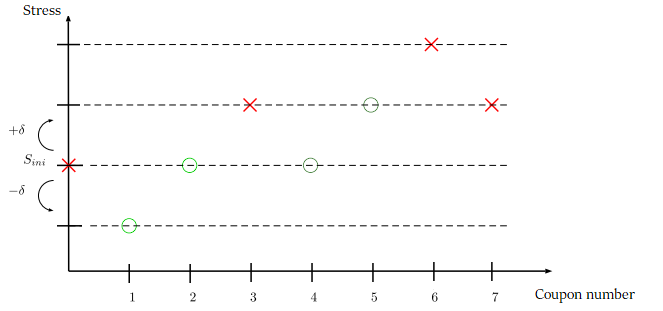}
\captionof{figure}{Staircase procedure}\label{staircase}
\end{center}

\textbf{Estimation}

After the $K$ trials, the parameter $\theta _{0}$ is estimated through
maximization of the likelihood, namely

\begin{equation}
\widehat \theta = \underset{\theta}{\text{argmax}}{\prod_{i=1}^K
\phi(s_i,\theta)^{y_i } (1-\phi(s_i,\theta))^{ (1-y_i) }}.
\end{equation}


\textbf{Numerical results}

The accuracy of the procedure has been evaluated on the two models presented below on a batch of 1000 replications,
each with $K=100.$

\smallskip
\emph{Exponential case}

Let $R\sim \mathcal{E}(\lambda)$ with $\lambda=0.2$. The input parameters are $S_{\text{ini}}=5 
$ and $\delta =15\in \left[ 0.5\times \frac{1}{\lambda ^{2}},2\times \frac{1%
}{\lambda ^{2}}\right] $.

As shown in Table \ref{stairexp}, the relative error pertaining to the parameter $\lambda $
is roughly $25\%$, although the input parameters are somehow optimal for the
method.\ The resulting relative error on the $10^{-3}$ quantile is $40\%.$
Indeed the parameter $\lambda $ is underestimated, which results in an
overestimation of the variance $1/\lambda ^{2}$ , which induces an
overestimation of the $10^{-3}$ quantile.

\begin{table}[]
\centering
\renewcommand{\arraystretch}{1.2} 
\begin{tabular}{|c|c|c|c|}
\hline
\multicolumn{4}{|c|}{\textbf{Relative error}} \\ 
\multicolumn{2}{|c|}{\textbf{On the parameter}} & \multicolumn{2}{c|}{%
\textbf{On $s_{\alpha}$}} \\ \hline\hline
Mean & Std & Mean & Std \\ \hline
-0.252 & 0.178 & 0.4064874 & 0.304 \\ \hline
\end{tabular}%
\caption{Results obtained using the \emph{Staircase} method through simulations under the exponential model.}
\label{stairexp}
\end{table}

\smallskip

\emph{Gaussian case }

We now choose $R\sim \mathcal{N}(\mu,\sigma)$ with $\mu=60$ and $\sigma=10$. The value of $S_{\text{ini}}$ is set
to the expectation and $\delta =7$ belongs to the interval $\left[ \frac{%
\sigma }{2},2\sigma \right] .$ The same procedure as above is performed and
yields the results in Table \ref{stairgaus}.
\begin{table}[tbp]
\centering
\renewcommand{\arraystretch}{1.2} 
\begin{tabular}{|c|c|c|c|c|c|}
\hline
\multicolumn{6}{|c|}{\textbf{Relative error}} \\ 
\multicolumn{2}{|c|}{\textbf{On $\mu$}} & \multicolumn{2}{c|}{\textbf{On $%
\sigma$}} & \multicolumn{2}{c|}{\textbf{On $s_{\alpha}$}} \\ 
\hline\hline
Mean & Std & Mean & Std & Mean & Std \\ \hline
-0.059 & 0.034 & 1.544 & 0.903 & -1.753 & 0.983 \\ \hline
\end{tabular}%
\caption{Results obtained using the \emph{Staircase} method through simulations under the Gaussian model.}
\label{stairgaus}
\end{table}

The expectation of $R$ is recovered rather accurately, whereas the
estimation of the standard deviation suffers a loss in accuracy, which in
turn yields a relative error of 180 \% \ on the $10^{-3}$ quantile.

\smallskip
\textbf{Drawback of the Staircase method}

A major advantage of the Staircase lies in the fact that the number
of trials to be performed in order to get a reasonable estimator of the mean is small. However, as shown by the simulations, this method is not adequate for the estimation of
extreme quantiles. Indeed, the latter follows from an extrapolation based on estimated
parameters, which furthermore may suffer of bias. Also, reparametrization
of the distribution making use of the theoretical extreme quantile would not
help, since the estimator would inherit of a large lack of accuracy.

\subsubsection{The Continuous Reassesment Method (CRM)}

\textbf{General principle}

The CRM (O'Quigley, Pepe and Fisher, 1990\cite{quigley})  has been designed for clinical trials and aims at the
estimation of $q_{\alpha }$ among $J$ stress levels $s_{1},...,s_{J}$, when $\alpha$ is of order $20\%$.

Denote $\mathbb{P}(R\leq s)=\psi (s,\beta _{0})$. The estimator of $q_{\alpha }$ is 
\[
s^{\ast }:=\underset{s_j \in \{s_{1},...,s_{J}\}}{\text{arginf}}{|\psi
(s_{j},\beta _{0})-\alpha |}.
\]%
This optimization is performed iteratively and $K$ trials are performed at each iteration.

Start with an initial estimator $\widehat{\beta _{1}}$ of $\beta _{0}$, for
example through a Bayesian choice as proposed in \cite{quigley}. Define 
\[
s_{1}^{\ast }:=\underset{s_j \in \{s_{1},...,s_{J}\}}{\text{arginf}}{|\psi
(s_{j},\widehat{\beta _{1}})-\alpha |}.
\]

Every iteration follows a two-step procedure:

\textbf{Step 1.} Perform $J$ trials under $\psi (.,\beta _{0})$, say $%
R_{1,1},..,R_{1,J}$ and observe only their value under threshold, say $%
Y_{1,j}:={\Large 1}_{R_{1,j}<s_{1}^{\ast }},1\leq j\leq J.$

\textbf{Step i. }Iteration $i$ consists in two steps :

\begin{itemize}
\item[--]  Firstly an estimate $\widehat{\beta _{i}}$ of $\beta _{0}$ is
produced on the basis of the information beared by the trials performed in all the preceding iterations through
maximum likelihood under $\psi (.,\beta _{0})$ (or by maximizing the posterior distribution of the parameter).

\item[--]\[
s_{i}^{\ast }:=\underset{s_j\in \{s_{1},...,s_{J}\}}{\text{arginf}}{|\psi
(s,}\widehat{{\beta _{i}}}{)-\alpha |};
\]

This stress level $s_{i}^{\ast }$ is the one under which the next $K$ trials $Y_{i,1},\dots,Y_{i,K}$ will be performed in the Bernoulli scheme $\mathcal{B}\left(\psi (s_{i}^{\ast },\beta _{0})\right)$.
\end{itemize}

The stopping rule depends on the context (maximum number of trials or
stabilization of the results).

Note that the bayesian inference is useful in the cases where there is no diversity in the observations at some iterations of the procedure, i.e when, at a given level of test $s_i^*$, only failures or survivals are observed.

\medskip

\textbf{Application to fatigue data}

The application to the estimation of the minimal allowable stress is treated in a bayesian setting. We do not directly put a prior on the parameter $\beta_0$, but rather on the probability of failure. We consider a prior information of the form: \emph{at a given stress level $s$, we can expect $k$ failures out of $n$ trials.} Denote $\pi_s$ the prior indexed on the stress level $s$. $\pi_s$ models the failure probability at level $s$ and has a Beta distribution given by
\begin{equation}\label{priorP}
\pi_{s} \sim \beta(k,n-k+1).
\end{equation}

Let $R$ follow an exponential distribution: $\forall s \ge 0,  \psi(s,\beta_0) = p_s = 1 - \exp(- \beta_0 s)$.

It follows $ \forall s,~ \beta_0 = - \frac{1}{s}\log(1- p_s)$.

Define the random variable  $\Lambda_s =  - \frac{1}{s}\log(1-\pi_{s})$ which, by definition of $\pi_s$, is distributed as an k-order statistic of a uniform distribution $U_{k,n}$.

The estimation procedure of the CRM is obtained as follows:

\textbf{Step 1. }Compute an initial estimator of the parameter
\[
\Lambda_{s} =  \frac{1}{L} \sum_{l=1}^L - \frac{1}{s}\log(1-\pi_{s}^{l} )
\]
with $\pi_{s}^l \sim \beta(k,n-k+1), 1\le l\le L$. Define
\[
s_{1}^{\ast }:=\underset{s_j\in \{s_{1},...,s_{J}\}}{\text{arginf}}{|( 1 - \exp(- \Lambda_s s_j))-\alpha |}.
\]
and perform $J$ trials at level $s_1^\ast$. Denote the observations $Y_{1,j}:={\Large 1}_{R_{1,j}<s_{1}^{\ast }},1\leq j\leq J.$

\textbf{Step i. }At iteration $i$, compute the posterior distribution of the parameter:

\begin{equation}
\pi^*_{s_i} \sim \beta \left(k + \sum_{l=1}^{i}\sum_{j=1}^{J}Y_{l,j}~,~ n + (J\times i) -(k + \sum_{l=1}^{i}\sum_{j=1}^{J}Y_{l,j}) +1 \right)
\end{equation}
The above distribution also corresponds an order statistic of the uniform distribution $U_{k + \sum_{l=1}^{i}\sum_{j=1}^{J}Y_{l,j}~,~n + (J\times i) }$. We  then obtain an estimate $\Lambda_{s_1^\ast}$.

The next stress level $s_{i+1}^{\ast }$ to be tested in the procedure is then given by
\[
s_{i+1}^{\ast }:=\underset{s_j\in \{s_{1},...,s_{J}\}}{\text{arginf}}{|( 1 - \exp(- \Lambda_{s_1^\ast} s_j))-\alpha |}.
\]

\smallskip
\textbf{Numerical simulation for the CRM}

Under the exponential model with parameter $\lambda =0.2$ and through $N=10$
iterations of the procedure, and $J=10$, with equally distributed thresholds 
$s_{1},..,s_{J}$ , and performing $K=50$ trials at each iteration, the
results in Table \ref{CRMexp} are obtained.

\begin{table}[]
\centering
\renewcommand{\arraystretch}{1.2} 
\begin{tabular}{|c|c|c|c|}
\hline
\multicolumn{4}{|c|}{\textbf{Relative error}} \\ 
\multicolumn{2}{|c|}{\textbf{On the $0.1-$quantile}} & 
\multicolumn{2}{c|}{\textbf{On the $10^{-3}-$quantile}} \\ 
\hline\hline
Mean & Std & Mean & Std \\ \hline
0.129 & 0.48 & -0.799 & 0.606 \\ \hline
\end{tabular}%
\caption{Results obtained through \emph{CRM} on simulations for the exponential model}
\label{CRMexp}
\end{table}

\begin{figure}[h]
\begin{center}
\includegraphics[scale=0.6]{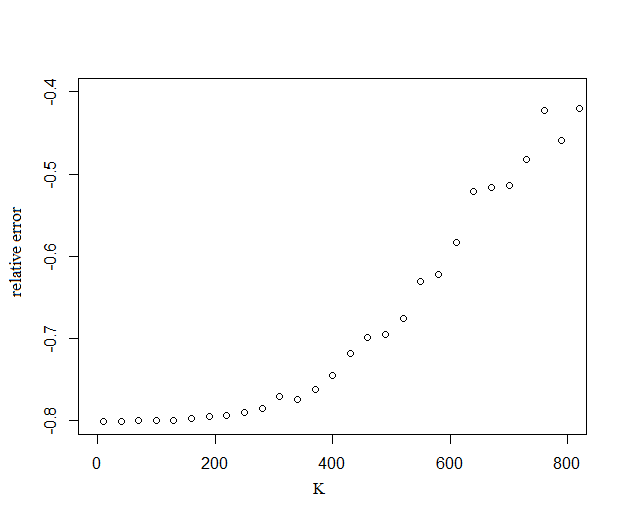}
\end{center}
\caption{Relative error on the  $10^{-3}$-quantile with respect to the number of trials for each stress level}
\label{crm}
\end{figure}

The $10^{-3}-$quantile is poorly estimated on a fairly simple model. Indeed for thresholds close to
the expected quantile, nearly no failure is observed. So, for acceptable $K$, 
the method is not valid; figure \ref{crm} shows the increase of accuracy
with respect to $K.$

Both the Staircase and the CRM have the same drawback in the context of
extreme quantile estimation, since the former targets the central tendency of the variable of interest and the latter aims at the estimation of quantiles
of order 0.2 or so, far from the target $\alpha =10^{-3}$. Therefore, we propose an original procedure designed for the estimation of extreme quantiles under binary information.

\section{A new design for the estimation of extreme quantiles}\label{Splitting}

\subsection{Splitting}
The design we propose is directly inspired by the general principle of Splitting methods used in the domain of rare events simulation and introduced by Kahn and Harris (1951 \cite{Kahn1951}). 

The idea is to overcome the difficulty of targeting an extreme event by decomposing the initial problem into a sequence of less complex estimation problem. This is enabled by the splitting methodology which decompose a small probability into the product of higher order probabilities.

Denote $\mathbb{P}$ the distribution of the r.v. $R$. The event $\{ R\le s_\alpha \}$ can be expressed as the intersection of inclusive events\: for $s_{\alpha }=s_{m}<s_{m-1}<...<s_{1}$ it holds: 
\[
\{R\leq s_{\alpha }\}=\{R\leq s_{m}\}\subset \dots \subset \{R\leq s_{1}\}.
\]

It follows that

\begin{equation}
\mathbb{P}(R\leq s_{\alpha })=\mathbb{P}(R\leq
s_{1})\prod_{j=1}^{m-1}\mathbb{P}(R\leq s_{j+1}\mid R\leq s_{j})
\label{Prod}
\end{equation}%
\label{split}

The thresholds $(s_{j})_{j=1,\dots ,m}$ should be chosen such that all $%
\mathbb{P}(R\leq s_{j+1}\mid R\leq s_{j})_{j=1,\dots ,m}$  be of order  $p=0.2$
or 0.3, in such a way that $\left\{ R\leq s_{j+1}\right\} $ is observed in
experiments performed under the conditional distribution of $R$ given $%
\left\{ R\leq s_{j}\right\} $, and in a way which makes $\alpha $ recoverable
by a rather small number of such probabilities $\mathbb{P}(R\leq s_{j+1}\mid
R\leq s_{j})$ making use of \eqref{Prod}.

From the formal decomposition in \eqref{Prod}, a practical experimental scheme can be deduced. Its form is given in algorithm \ref{Split}.

\begin{algorithm}[H]
\floatname{algorithm}{Procedure}
\caption{Splitting procedure}
\begin{algorithmic}\label{Split}
\STATE \textbf{Initialization}

\STATE Fix 
\begin{itemize}
$\left.  \text{\parbox{0.6\linewidth}{
\item[$\bullet$] the number $m$ of iterations to be performed (and of levels to be tested);
\item[$\bullet$] the level of conditional probabilities $p$  (laying between 20 and 30 \%);
}} \right \}$  ~~~ such that $p^m \approx \alpha$
\medskip 
\item[$\bullet$] the first tested level $s_1$ (ideally the $p-$quantile of the distribution of $R$);
\item[$\bullet$] the number $K$ of trials to be performed at each iteration.
\end{itemize}

\smallskip
\STATE \textbf{First step}
\begin{itemize}
\item[$\bullet$] $K$ trials are performed at level $s_1$. The observations are the indicators of failure $Y_{1,1},\dots,Y_{1,K}$, where $Y_{1,i} = \mathds{1}(R_{1,i}<s_1)$ of distribution $\mathcal{B}\left( \mathbb{P}(R \le s_1)\right)$.
\item[$\bullet$]  Determination of $s_2$, $p-$quantile of the truncated distribution $R \mid R \le s_1$.
\end{itemize}

\medskip
\STATE \textbf{Iteration $j=2$ to $m$}
\begin{itemize}
\item[$\bullet$] $K$ trials are performed at level $s_j$ under the truncated distribution of $R \mid R \le s_{j-1}$ resulting to observations $Y_{j,1},\dots,Y_{j,k} \sim \mathcal{B}\left( \mathbb{P}(R \le s_j\mid R \le s_{j-1})\right)$.
\item[$\bullet$]  Determination of $s_{j+1}$, the $p-$quantile of $R \mid R \le s_{j}$. 
\end{itemize}

\smallskip
\STATE The last estimated quantile $s_m$ provides the estimate of $s_\alpha$.

\end{algorithmic}
\end{algorithm}

\subsection{Sampling under the conditional probability}\label{operational_procedure}

In practice batches of specimen are put under trial, each of them with a
decreasing strength; this allows to target the tail of the distribution $%
\mathbb{P}$ iteratively.

\begin{center}
\includegraphics[scale=0.4]{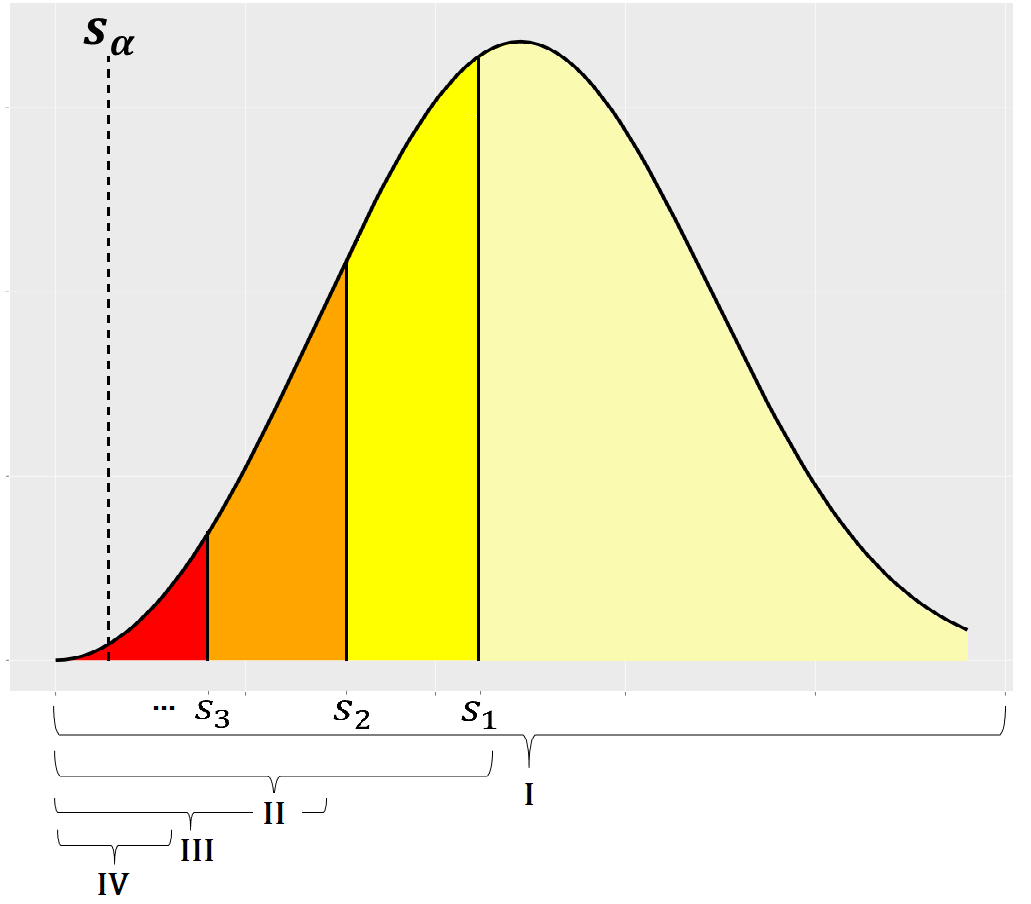} 
\captionof{figure}{Sampling under the strengh density at $n_0$ cycles}\label{fatigue}
\end{center}

In other words, in the first step, points are sampled in zone (I). Then in the following step, only specimen with strength in zone II
are considered, and so on. In the final step, the specimen are sampled in zone IV. At level $s_{m}$, they have a very small probability to fail before $n_{0}$ cycles under $\mathbb{P}$, however under their own law of failure, which is $\mathbb{P}(\mathbf{.}\mid R\leq s_{m-1})$, they
 have a probability of failure of order 0.2.

\bigskip 

In practice, sampling in the tail of the distribution is achieved by introducing flaws in the batches of specimens. The idea is that the strength of the material varies inversely with respect to the size of the incorporated flaws. The flaws are spherical and located inside the specimen (not on its surface). Thus, as the procedure moves on, the trials are performed on samples of materials incorporating flaws of greater diameter. This procedure is based on the hypothesis that there is a correspondence between the strength of the material with flaw of diameter $\theta$ and the truncated strength of this same material without flaw under level of stress $s^*$, i.e. we assume that noting $R_{\theta }$ the strength of the specimen with flaw of size $\theta $, it holds that there exists $s^*$ such that 
\[
\mathcal{L}(R_{\theta })\approx \mathcal{L}(R\mid R\leq s^*).
\]

Before launching a validation campaign for this procedure, a batch of 27 specimen has been machined including spherical defects whose sizes vary between 0 and 1.8mm (see Figure  \ref{defautseprouvettes}). These first trials aim at estimating the decreasing relation between mean allowable stress and defects diameter $\theta$. This preliminary study enabled to draw the abatement fatigue curve as a function of $\theta$, as shown in Figure \ref{abattement}.

\begin{center}
\includegraphics[scale=0.4]{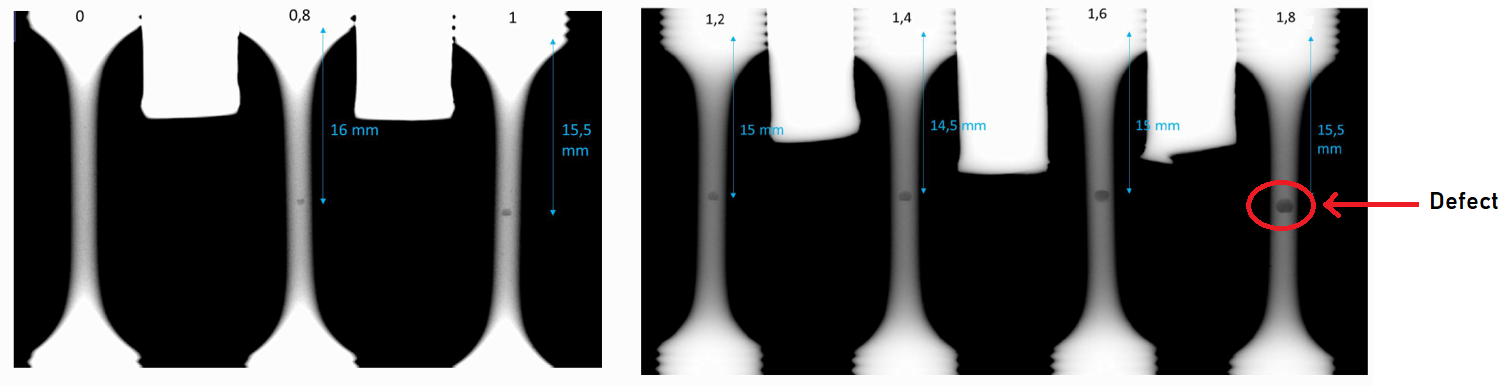}
\captionof{figure}{Coupons incorporating spherical defects of size varying from 0 mm (on the left) to 1.8 mm (on the right)}\label{defautseprouvettes}
\end{center}
\bigskip

\begin{center}
\includegraphics[scale=0.3]{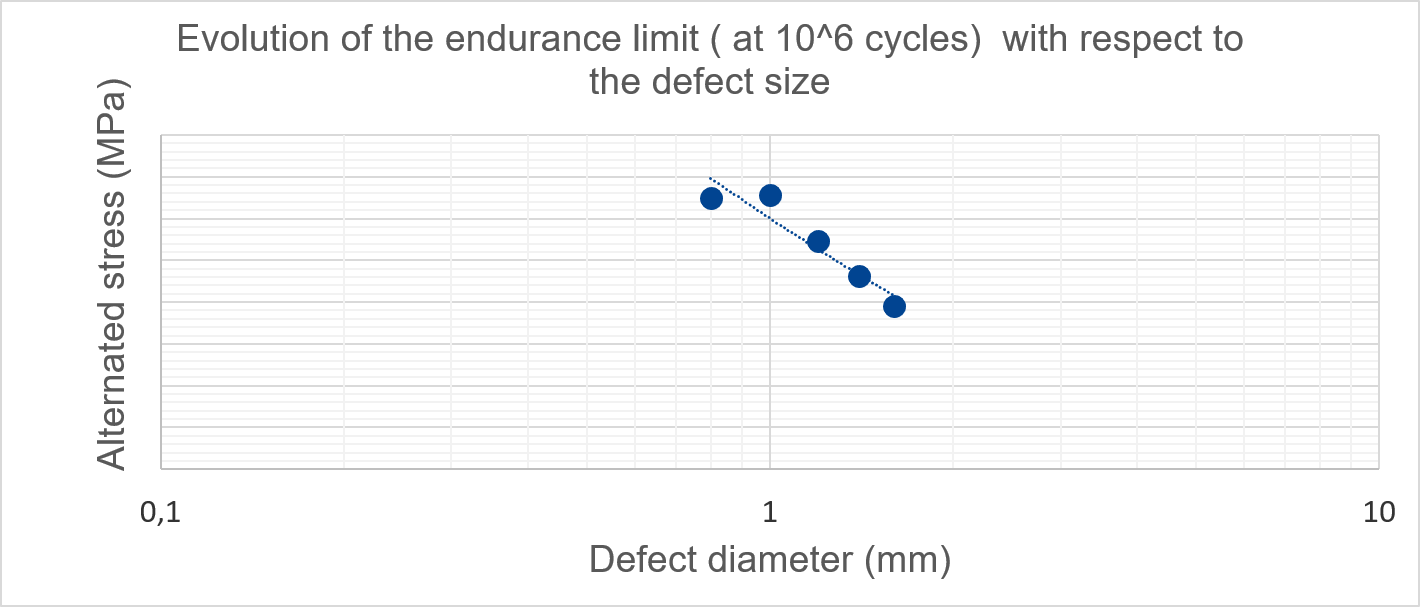}
\captionof{figure}{Mean allowable stress with respect to the defect size}\label{abattement}
\end{center}
\bigskip

Results in  Figure \ref{abattement} will be used during the splitting procedure to select the diameter $\theta$ to be incorporated in the batch of specimens tested at the current iteration as reflecting the sub-population of material of smaller resistance.

\subsection{Modeling the distribution of the strength, Pareto model}\label{GPD_model}

The events under consideration have small probability under $\mathbb{P}.$
By (\ref{Prod}) we are led to consider the limit behavior of conditional
distributions under smaller and smaller thresholds, for which we make use of
classical approximations due to Balkema and de Haan (1974\cite{bal}) which stands as follows,
firstly in the commonly known setting of exceedances over increasing
thresholds. Denote $\widetilde{R}:=1/R$.

\begin{theorem}
\label{Thm de Haan}For $\widetilde{R}$ of distribution $F$ belonging to the maximum domain of attraction of an extreme value distribution with tail index $c$, i.e. $F\in MDA(c)$, it holds that: There exists $a=a(s)>0$, such that:
\[ \lim\limits_{s \rightarrow \infty}\sup_{0\le x< \infty} 
\left\lvert 
\frac{1 - F\left(x + s\right)}{1 - F\left( s \right)} -  \left(1 - G_{(c,a}(x)\right)
\right\rvert = 0
\]
where  $G_{(c,a)}$ is defined through $~$%
\begin{equation*}
G_{(c,a)}(x)=1-\exp \left\{ -\int_{0}^{\frac{x}{a}}\left[ (1+ct)_{+}\right]
^{-1}dt\right\} 
\end{equation*}
where  $a>0$ and $c\in \mathbb{R}$.
\end{theorem}

The distribution $G$ is the Generalized Pareto distribution $GPD(c,a)$ is defined explicitly through%
\[
1-G(x)=\left\{ 
\begin{array}{l}
(1+\frac{c}{a}x)^{-1/c}\text{ when }c\neq 0 \\ 
\exp (-\frac{x}{a})\text{ when }c=0%
\end{array}%
\right. 
\]%
where $x\geq 0$ for $c\geq 0$ and\ $0\leq x\leq -\frac{a}{c}$ if $c<0.$

Generalized Pareto distributions enjoy invariance through threshold
conditioning, an important property for our sake. Indeed it holds, for $%
\widetilde{R}\sim GDP(c,a)$ and $x>s$,

\begin{equation}
\mathbb{P}\left( \widetilde{R}>x\mid \widetilde{R}>s\right) =\left( 1+\frac{%
c(x-s)}{a+cs}\right) ^{-1/c}  \label{r}
\end{equation}%
\bigskip 
We therefore state:
\begin{proposition}
\label{Prop stability GPD}When $\widetilde{R}\sim GPD(c,a)$ then, given $%
\left( \widetilde{R}>s\right) $, the r.v. $\widetilde{R}-s$ follows a $%
GPD(c,a+cs)$.
\end{proposition}

The GPD's are on the one hand stable under thresholding and on the other appear as the limit distribution for thresholding operations. This chain of arguments is quite usual in statistics, motivating the recourse to the ubiquous normal or stable laws for additive models. This plays in favor of GPD's as modelling the distribution of $\widetilde{R}$ for excess probability inference. Due to the lack of memory property, the
exponential distribution which appears as a possible limit distribution for
excess probabilities in Theorem \ref{Thm de Haan} do not qualify for
modelling.  
Moreover since we handle variables $R$ which can approach $0$ arbitrarily (i.e.
unbounded $\widetilde{R}$) the parameter $c$ is assumed positive.

Turning to the context of the minimal admissible constraint, we make use of
the r.v. $R=1/\widetilde{R}$ and proceed to the corresponding change of
variable.

When $c>0$, the distribution function of the r.v. $R$ writes for nonnegative $x$:
\begin{equation}\label{GPD 1}
F_{c,a}(x)=(1+\frac{c}{ax})^{-1/c}.
\end{equation}

For $0<x<u$, the conditional distribution of $R$ given $\left\{ R<u\right\} $
is 
\[
\mathbb{P}(R<x\mid R<u)=\left( 1-\frac{c(\frac{1}{x}-\frac{1}{u})}{a+\frac{c%
}{u}}\right) ^{-1/c}
\]%

which proves that the distribution of $R$ is stable under threshold
conditioning with parameter $\left( a_{u},c\right) $ with 
\begin{equation}
a_{u}=a+\frac{c}{u}.  \label{transition param a}
\end{equation}%
In practice\ at each step $j$ in the procedure the stress level $s_{j}$
equals the corresponding threshold $1/\widetilde{s}_{j}$ , a right quantile
of the conditional distribution of $\widetilde{R\text{ }}$ given $\left\{ 
\widetilde{R}>\widetilde{s}_{j-1}\right\} $. Therefore the observations take
the form $Y_{i}=\mathds{1}_{R_{i}<s_{j-1}}=\mathds{1}_{\widetilde{R}_{i}>%
\widetilde{s}_{j-1}},~~i=1,\dots ,K_{j}$.

A convenient feature of model (\ref{GPD 1}) lies in the fact that the
conditional distributions are completely determined by the initial
distribution of $R$ , therefore by $a\ $\ and $c.$ The parameters $a_{j}$ of
the conditional distributions are determined from these initial parameters
and by the corresponding stress level $s_{j};$ see (\ref{transition param a}).

\subsection{Notations}

The distribution function of the r.v. $\widetilde{R}$ is a $%
GPD(c_{T},a_{T})$ of distrubution function $G_{(c_{T},a_{T})}.$ Note $%
\overline{G}_{(c_{T},a_{T})}=1-G_{(c_{T},a_{T})}.$

Our proposal relies on iterations. We make use of a set of thresholds $(%
\widetilde{s}_{1},...,\widetilde{s}_{m})$ and define for any $j\in
\{1,...,m\}$ 

\[
G_{(c_{j},a_{j})}(x - \widetilde{s}_{j})=\mathbb{P(}\left. \widetilde{R}>x\right\vert \widetilde{%
R}>\widetilde{s}_{j})
\]

with $c_{j}=c_{T}$ and $a_{j}=a_{T}+c_{T}\widetilde{s}_{j}$ where we used (%
\ref{r}).

At iteration $j$, denote $(\widehat{c},\widehat{a})_{j}$ the estimators of $%
(c_{j},a_{j})$.Therefore $1-G_{(\widehat{c},\widehat{a})_{j}}(x - \widetilde{s}_{j})$ estimates $%
\mathbb{P(}\left. \widetilde{R}>x\right\vert \widetilde{R}>\widetilde{s}_{j})
$. Clearly, estimators of $(c_{T},a_{T})$ can be recovered from $(\widehat{c}%
,\widehat{a})_{j}$ through $\widehat{c}_{T}=\widehat{c}$ and $\widehat{a}%
_{T}=\widehat{a}-\widehat{c}~\widetilde{s}_{j}.$

\subsection{Sequential design for the extreme quantile estimation}\label{proc}

Fix $m$ and $p$ , where $m$ denotes the number of stress levels under which
the trials will be performed, and $p$ is such that $p^m=\alpha .$

Set a first level of stress, say $s_{1}$ large enough (i.e. $\widetilde{s}_{1}=1/s_{1}$ small enough) 
so that $p_1 = \mathbb{P}(R<s_{1})$ is large enough and perform trials at this level. The optimal
value of $s_{1}$ should satisfy $p_1=p$, which cannot be
secured. This choice is based on expert advice.

 Turn to $\widetilde{R}:=1/R$. Estimate $c_{T}$ and $a_{T}$, for
the GPD $\left( c_{T},a_{T}\right) $ model describing $\widetilde{R}$, say $(%
\widehat{c},\widehat{a})_1$, based on the observations above $\widetilde{s}_{1}$
(note that under $s_{1}$ the outcomes of $R$ are easy to obtain, since the
specimen is tested under medium stress).

Define 
\[
\widetilde{s}_{2}:=\sup \left\{ s:\overline{G}_{(\widehat{c},\widehat{a}%
)_{1}}\left( s - \widetilde s_1\right) <p\right\} 
\]%
the $(1-p)-$quantile of $G_{(\widehat{c},\widehat{a})_{1}}.$ $\widetilde s_2$ is the level of stress to be tested at the following iteration.

Iterating from step $j=2$ to $m-1$, perform $K$ trials under $G_{(c_1,a_1)}
$ say $\widetilde{R}_{j,1},..,\widetilde{R}_{j,K}$ and consider the
observable variables $Y_{j,i}:={\Large 1}_{\widetilde{R}_{j,i}>\widetilde{s}_{j}}$.
 Therefore the $K$ iid replications $Y_{j,1},..,Y_{j,K}$ follow a
Bernoulli $\mathcal{B}(\overline{G}_{({c}_{j-1},{a}_{j-1})}\left( \widetilde s_{j} - \widetilde s_{j-1}\right) )$, where $\widetilde s_j$ has been determined at the previous step of the procedure. Estimate $(c_{j},a_{j})$ in the
resulting Bernoulli scheme, say $(\widehat{c},\widehat{a})_{j}$. Then define 
\[
\begin{split}
\widetilde s_{j+1}&:=\sup \left\{ s:\overline{G}_{(\widehat{c},\widehat{a}%
)_{j}}\left( s - \widetilde s_j\right) <p\right\} \\
&=G_{\left( \widehat{c},\widehat{a}\right)
_{j}}^{-1}(1-p)+\widetilde{s}_{j},
\end{split}
\] 
which is the $(1-p)-$quantile of the estimated conditional distribution of $\widetilde{R}$ given $\{ \widetilde{R}>\widetilde{s_{j}}\}$, i.e. $G_{(%
\widehat{c},\widehat{a})_{j}}$, and the next level to be tested.

In practice a conservative choice for $m$ is given by $m=\left\lceil \frac{%
log\alpha }{logp}\right\rceil $, where $\lceil. \rceil $ denotes the ceiling function. This implies that the attained probability 
$\widetilde{\alpha}$ is less than or equal to $\alpha.$

The $m$ stress levels $\widetilde{s}_{1}<\widetilde{s}_{1}<\dots <%
\widetilde{s}_{m}=\widetilde{q}_{1-\alpha}$ satisfy 
\[
\begin{split}
\widetilde{\alpha }& =\overline{G}(\widetilde{s}_{1})\prod_{j=1}^{m-1}\overline{G}_{\left( \widehat{c},\widehat{a}\right)
_{j}}(\widetilde{s}_{j+1} - \widetilde s_j) \\
& ={p}_{1}p^{m-1}
\end{split}%
\]

Finally by its very definition $\widetilde{s}_{m}$ is a proxy of $\widetilde{%
q}_{1-\alpha }.$

Although quite simple in its definition, this method bears a number of
drawbacks, mainly in the definition of $\left( \widehat{c},\widehat{a}%
\right) _{j}.$ The next section addresses this question.


\section{Sequential enhanced design in the Pareto model}\label{EstimationProc}

In this section we focus on the estimation of the parameters $\left(
c_{T},a_{T}\right) $ in the $GPD(c_{T},a_{T})$ distribution of $\widetilde{R}%
.$ One of the main difficulties lies in the fact that the available
information does not consist of replications of the r.v. $\widetilde{R}$
under the current conditional distribution $G_{(c_{j},a_{j})}$ of $%
\widetilde{R}$ given $\left( \widetilde{R}>\widetilde{s_{j}}\right) $ but
merely on very downgraded functions of those. 

At step $j$ we are given $G_{(\widehat{c},\widehat{a})_{j}}$ and define $%
\widetilde{s}_{j+1}$ as its $\left( 1-p\right) -$quantile. Simulating $K$
r.v. $\widetilde{R}_{j,i}$ with distribution $G_{(c_{j},a_{j})}$, the observable
outcomes are the Bernoulli ($p$) r.v.'s $Y_{j,i}:=1_{\widetilde{R}_{j,i}>%
\widetilde{s}_{j+1}}.$ This loss of information with respect to the $%
\widetilde{R}_{j,i}$ 's makes the estimation step for the coefficients $(%
\widehat{c},\widehat{a})_{j+1}$ quite complex; indeed $(\widehat{c},\widehat{%
a})_{j+1}$ is obtained through the $Y_{j,i}$'s, $1\leq i\leq K$.

\begin{figure}[h]
\begin{center}
\includegraphics[scale=0.4]{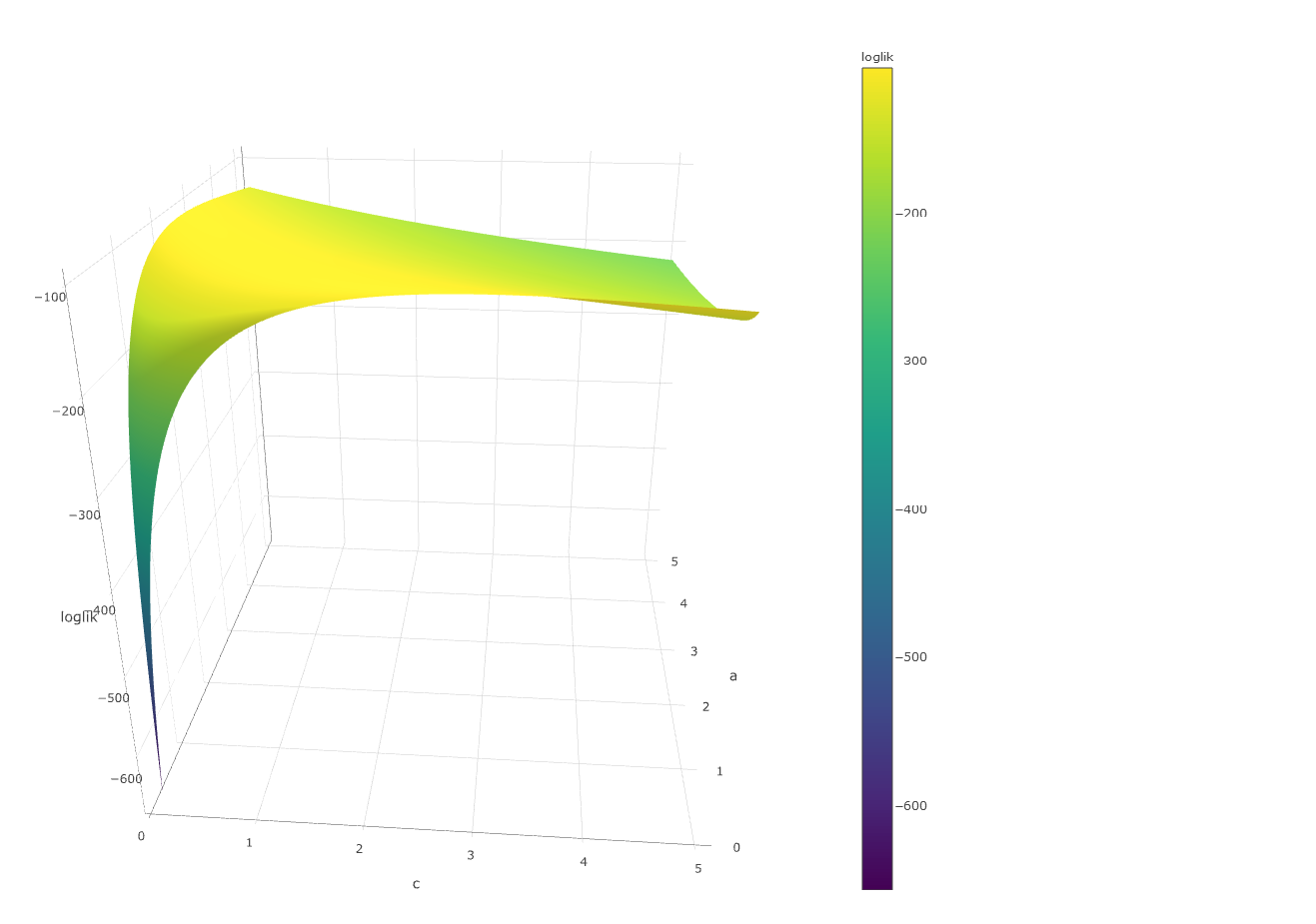}
\end{center}
\caption{Log-likelihood of the Pareto model with binary data}
\label{loglik}
\end{figure}

It is of interest to analyze the results obtained through standard Maximum
Likelihood Estimation of $(\widehat{c},\widehat{a})_{j+1}.$ The quantile $%
\widetilde{q}_{1-\alpha }$ is loosely estimated for small $\alpha $; as
measured on 1000 simulation runs, large standard deviation of $\widehat{%
\widetilde{q}}_{1-\alpha }$ is due to poor estimation of the iterative
parameters $(\widehat{c},\widehat{a})_{j+1}.$ We have simulated $n=200$
realizations of r.v.'s $Y_{i}$ with common Bernoulli distribution with
parameter $\overline{G}_{\left( c_{T},a_{T}\right) }(\widetilde{s}_{1}).$
Figure \ref{loglik} shows the log likelihood function of this sample as the
parameter of the Bernoulli $\overline{G}_{\left( c^{\prime },a^{\prime
}\right) }(\widetilde{s_{0}})$ varies according to $\left( c^{\prime
},a^{\prime }\right) .$ As expected this function is nearly flat in a very
large range of $\left( c^{\prime },a^{\prime }\right) .$

This explains the poor results in Table \ref{procQuantm} obtained through the Splitting procedure when the parameters at each step 
are estimated by maximum likelihood, especially in terms of dispersion of the estimations. Moreover, the accuracy  of the
estimator of $\widetilde{q}_{1-\alpha }$ quickly decreases  with the number $K$ of replications $Y_{j,i}$, $1\leq i\leq K$. 

\begin{figure}[h]
\begin{center}
\includegraphics[scale=0.6]{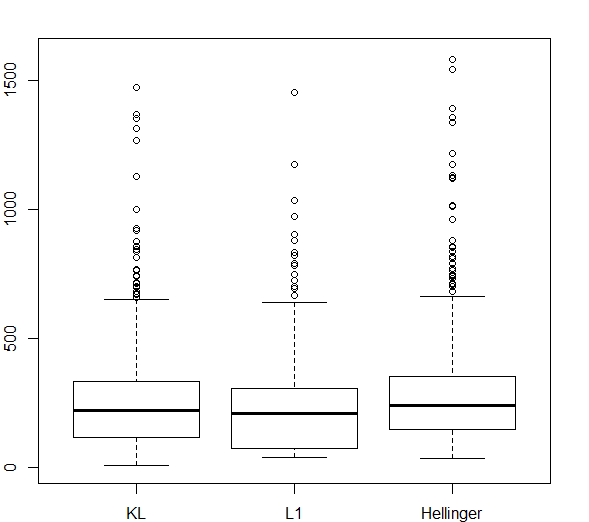}
\end{center}
\caption{Estimations of the $\protect\alpha-$quantile based on the 
Kullback-Leibler, L1 distance and Hellinger distance criterion}
\label{disp_quant}
\end{figure}

Changing the estimation criterion by some alternative method does not
improve significantly; Figure \ref{disp_quant} shows the distribution of the
resulting estimators of $\widetilde{q}_{1-\alpha }$ for various estimation
methods (minimum Kullback Leibler, minimum Hellinger and minimum L1
distances - see their definitions in Appendix \ref{div}) of $\left( c_{T},a_{T}\right).$

This motivates the need for an enhanced estimation procedure.

\begin{table}[]
\centering%
\begin{tabular}{c|c|c|c|c|c}
\hline
Minimum & Q25 & Q50 & Mean & Q75 & Maximum \\ \hline
67.07 & 226.50 & 327.40 & 441.60 & 498.90 & 10 320.00 \\ \hline
\end{tabular}%
\caption{Estimation of the $(1-\alpha)-$quantile, $\widetilde s_{\alpha}%
=469.103$, through procedure \protect\ref{proc} with $K=50$}
\label{procQuant}
\end{table}

\begin{table}[]
\centering%
\begin{tabular}{|c|c|c||c|c|}
\hline
& \multicolumn{2}{c||}{$\widetilde s_m$ for $K=30$} & \multicolumn{2}{c|}{$\widetilde s_m$ for $%
K=50$} \\ 
\textbf{$\widetilde s_{\alpha}$} & Mean & Std & Mean & Std \\ \hline
469.103 & 1 276.00 & 12 576.98 & 441.643 & 562.757 \\ \hline
\end{tabular}%
\caption{Estimation of the $(1-\alpha)-$quantile, $\widetilde s_{\alpha}%
=469.103$, through procedure \protect\ref{proc}  for different
values of $K$}
\label{procQuantm}
\end{table}

\bigskip

\bigskip

\bigskip 

\subsection{An enhanced sequential criterion for estimation}\label{procImpML}

We consider an additional criterion which makes a peculiar use of the
iterative nature of the procedure. We will impose some control on the
stability of the estimators of the conditional quantiles through the
sequential procedure.

At iteration $j-1$, the sample $Y_{j-1,i}$ , $1\leq i\leq K$ has been
generated under $G_{(\widehat{c},\widehat{a})_{j-2}\text{ }}$and provides an
estimate of $p$ through 
\begin{equation}
\widehat{p}_{j-1}:=\frac{1}{K}\sum_{i=1}^{n}Y_{j-1,i}.  \label{p_ji}
\end{equation}%
The above $\widehat{p}_{j-1}$ estimates $\mathbb{P}\left( \widetilde{R}>%
\widetilde{s}_{j-1}\mid \widetilde{R}>\widetilde{s}_{j-2}\right) $
conditionally on $\widetilde{s}_{j-1}$ and $\widetilde{s}_{j-2}.$ We write
this latter expression $\mathbb{P}\left( \widetilde{R}>\widetilde{s}%
_{j-1}\mid \widetilde{R}>\widetilde{s}_{j-2}\right) $ as a function of the
parameters obtained at iteration $j$ , namely $(\widehat{c},\widehat{a})_{j}.
$The above r.v's $\ Y_{j-1,i}$ stem from variables $\widetilde{R}_{j-1,i}$
greater than $\ \widetilde{s}_{j-2}.$ At step $j,$ estimate then $\mathbb{P}%
\left( \widetilde{R}>\widetilde{s}_{j-1}\mid \widetilde{R}>\widetilde{s}%
_{j-2}\right) $ making use of $G_{(\widehat{c},\widehat{a})_{j}}.$ This backward
estimator writes 
\[
\frac{\overline{G}_{(\widehat{c},\widehat{a})_{j}}(\widetilde{s}_{j-1})}{%
\overline{G}_{(\widehat{c},\widehat{a})_{j}}(\widetilde{s}_{j-2})}=1-G_{(%
\widehat{c},\widehat{a})_{j}}(\widetilde{s}_{j-1}-\widetilde{s}_{j-2}).
\]%
The distance 
\begin{equation}
\left\vert \left( \overline{G}_{(\widehat{c},\widehat{a})_{j}}(\widetilde{s}%
_{j-1}-\widetilde{s}_{j-2})\right) -\widehat{p_{j-1}}\right\vert   \label{A}
\end{equation}%
should be small, since both $ \overline{G}_{(\widehat{c},\widehat{a}%
)_{j}}(\widetilde{s}_{j-1}-\widetilde{s}_{j-2}) $ and $\ \widehat{%
p}_{j-1}$ should approximate $p.$

Consider the distance between quantiles 
\begin{equation}
\left\vert (\widetilde{s}_{j-1}-\widetilde{s}_{j-2})-G_{(\widehat{c},%
\widehat{a})_{j}}^{-1}(1-\widehat{p}_{j-1})\right\vert .  \label{B}
\end{equation}

An estimate $(\widehat{c},\widehat{a})_{j}$ can be proposed as the minimizer
of the above expression for $(\widetilde{s}_{j-1}-\widetilde{s}_{j-2})$ for
all $j$. This backward estimation provides coherence with respect to the
unknown initial distribution $G_{\left( c_{T},a_{T}\right) }$. Would we have
started with a good guess $(\widehat{c},\widehat{a})=\left(
c_{T},a_{T}\right) $ then the successive $(\widehat{c},\widehat{a})_{j},\ 
\widetilde{s}_{j-1}$ etc would make (\ref{B}) small, since $\widetilde{s}%
_{j-1}$ (resp. $\widetilde{s}_{j-2}$) would estimate the $p-$conditional
quantile of $\mathbb{P}\left( \left. .\right\vert \widetilde{R}>\widetilde{s}%
_{j-2}\right) $ (resp. $\mathbb{P}\left( \left. .\right\vert \widetilde{R}>%
\widetilde{s}_{j-3}\right) $).\ 

It remains to argue on the set of plausible values where the quantity in (%
\ref{B}) should be minimized.

We suggest to consider a confidence region for the parameter $\left(
c_{T},a_{T}\right) .$ With $\widehat{p}_{j}$ defined in (\ref{p_ji}) and $%
\gamma \in \left( 0,1\right) $ define the $\gamma -$confidence region for $p$
by

\[
I_{\gamma }=\left[ \widehat{p}_{j}-z_{1-\gamma /2}\sqrt{\frac{\widehat{p}%
_{j}(1-\widehat{p}_{j})}{K-1}};\widehat{p}_{j}+z_{1-\gamma /2}\sqrt{\frac{%
\widehat{p}_{j}(1-\widehat{p}_{j})}{K-1}}\right] 
\]%
where $z_{\tau }$ is the $\tau -$quantile of the standard normal
distribution. Define 
\[
\mathcal{S}_{j}=\left\{ (c,a):\left( 1-G_{(c,a)}(\widetilde{s}_{j}-%
\widetilde{s}_{j-1})\right) \in I_{\gamma }\right\} .
\]%
Therefore $\mathcal{S}_{j}$ is a plausible set for $(\widehat{c}_{T},%
\widehat{a}_{T}).$

We summarize this discussion:

At iteration $j,$ the estimator of $\left( c_{T},a_{T}\right) $ is a
solution of the minimization problem 
\[
\min_{(c,a)\in \mathcal{S}_{j}}\left\vert (\widetilde{s}_{j-1}-\widetilde{s}%
_{j-2})-G_{(c,a+c\widetilde{s}_{j-2})}^{-1}(1-\widehat{p}_{j-1})\right\vert .
\]
The optimization method used is the Safip algorithm (Biret and Broniatowski, 2016 \cite{Biret})
As seen hereunder, this heuristics provides good performance.

\subsection{Simulation based numerical results\label{Subsection numGPD}}

This procedure has been applied in three cases. A case considered as
reference is $(c_{T},a_{T})=(1.5,1.5)$; secondly the case when $%
(c_{T},a_{T})=(0.8,1.5)$ describes a light tail with respect to the
reference.\ Thirdly, a case $(c_{T},a_{T})=(1.5,3)$ defines a distribution
with same tail index as the reference, but with a larger dispersion index.

Table \ref{parIC} shows that the estimation of $\widetilde{q}_{1-\alpha }$
deteriorates as the tail of the distribution gets heavier; also the
procedure underestimates $\widetilde{q}_{1-\alpha }.$ 
\begin{table}[tbp]
\centering
\renewcommand{\arraystretch}{1.2} 
\begin{tabular}{|c||c|c|}
\hline
\textbf{Parameters} & \multicolumn{2}{c|}{\textbf{Relative error on $%
\widetilde s_{\alpha}$}} \\ 
& Mean & Std \\ \hline\hline
$c=0.8$, $a_0=1.5$ and $\widetilde s_{\alpha}= 469.103$ & -0.222 & 0.554 \\ \hline
$c=1.5$, $a_0=1.5$ and $\widetilde s_{\alpha}=31621.777 $ & -0.504 & 0.720 \\ \hline
$c=1.5$, $a_0=3$ and $\widetilde s_{\alpha}=63243.550 $ & 0.310 & 0.590 \\ \hline
\end{tabular}%
\caption{Mean and std of relative errors on the $(1-\alpha)-$quantile of GPD calculated through 400 replicas of procedure \ref{procImpML}.}
\label{parIC}
\end{table}

\begin{figure}[]
\includegraphics[scale=0.45]{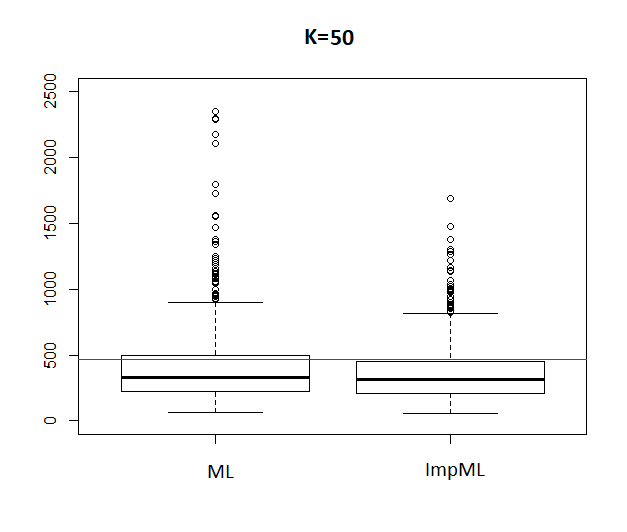} %
\includegraphics[scale=0.45]{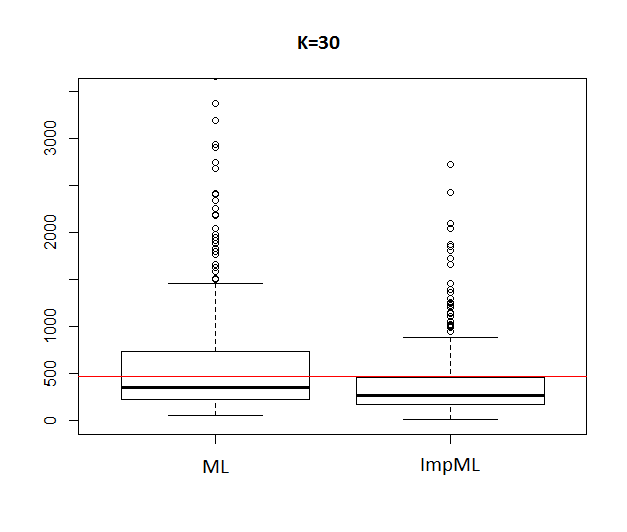} %
\includegraphics[scale=0.45]{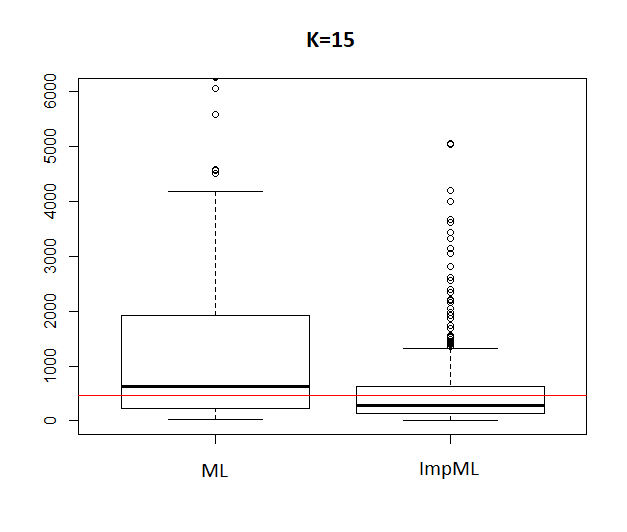}

{\small The red line stands stands for the real value of $s_\alpha$ }
\caption{Estimations of the $(1-\alpha)-$quantile of a $GPD(0.8,1.5)$ obtained by Maximum Likelihood and by the improved Maximum Likelihood method for different values of $K$.}
\label{compMVQm}
\end{figure}

Despite these drawbacks, we observe an improvement with respect to the simple Maximum Likelihood estimation; this is even more clear, when the tail of the distribution is heavy. Also, in contrast
with the ML estimation, the sensitivity with respect to the number $K$ of replications at each of the iterations plays in favor of this new method: As $K$ decreases, the gain with respect to Maximum Likelihood estimation increases notably, see Figure \ref{compMVQm}.


\begin{figure}[]\label{compMVQ}
\includegraphics[scale=0.45]{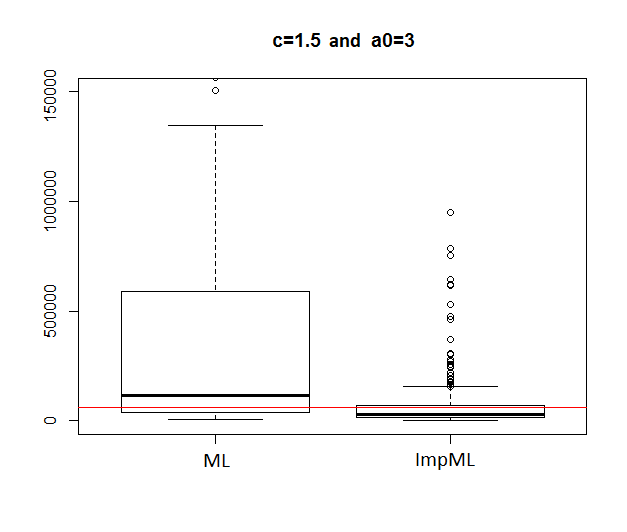} 
\includegraphics[scale=0.45]{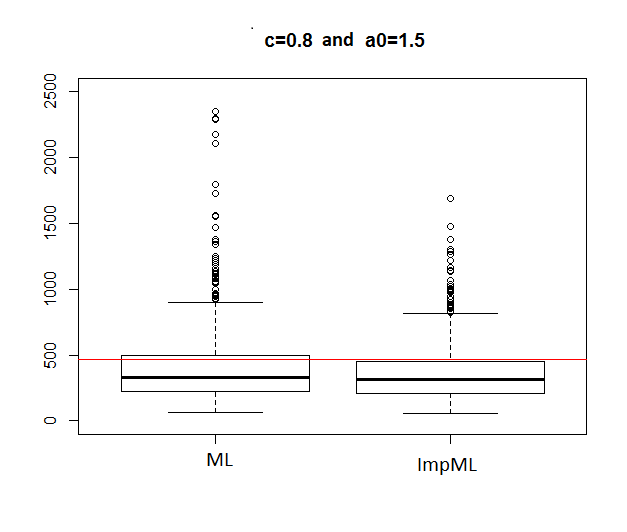} %
{\small The red line stands stands for the real value of $s_\alpha$ }
\caption{Estimations of the $(1-\alpha)-$quantile of two GPD obtained by Maximum Likelihood and by the improved Maximum Likelihood method}
\end{figure}

\subsection{Performance of the sequential estimation\label{subsection
comparaison de Valk GPD}}

As stated in chapter \ref{revLit}, there is to our knowledge no method dealing with similar question available in the
literature. 
Therefore we compare the results of our method, based on observed exceedances over thresholds, with the results
that could be obtained by classical extreme quantiles estimation methods assuming we have complete data at our disposal; those may
be seen as benchmarks for an upper bound of the performance of our method.

\subsubsection{Estimation of an extreme quantile based on complete data, de
Valk's estimator}

In order to provide an upper bound for the performance of the estimator, we
make use of the estimator proposed by De Valk and Cai (2016). This work aims at
the estimation of a quantile of order $p_{n}\in \lbrack n^{-\tau
_{1}};n^{-\tau _{2}}]$, with $\tau _{2}>\tau _{1}>1$ , where $n$ is the
sample size. This question is in accordance with the industrial context
which motivated the present paper. De Valk's proposal is a modified Hill
estimator adapted to log-Weibull tailed models. De Valk's estimator is
consistent, asymptotically normally distributed, but is biased for finite
sample size.
We briefly recall some of the hypotheses which set the context of de Valk's
approach.

Let $X_{1},\dots ,X_{n}$ be $n$ iid r.v's with distribution $F$, and denote $%
X_{k:n}$ the $k-$ order statistics. A tail regularity assumption is needed
in order to estimate a quantile with order greater than $1-$ $1/n.$

Denote $U(t)=F^{-1}\left( 1-1/t\right) $, and let the function $q$ be
defined by

\[
q(y)=U(e^{y})=F^{-1}\left( 1-e^{-y}\right) 
\]

for $y>0$. 

Assume that  
\begin{equation}
\lim\limits_{y\rightarrow \infty }~\frac{\log q(y\lambda )-\log q(y)}{g(y)}%
=h_{\theta }(\lambda )~~~\lambda >0  \label{logweibulltail}
\end{equation}%
where $g$ is a regularly varying function and 
\[
h_{\theta }(\lambda )=\left\{ 
\begin{array}{l}
\frac{\lambda ^{\theta }-1}{\theta }\text{ if }\theta \neq 0 \\ 
\log \lambda \text{ if }\theta =0%
\end{array}%
\right. 
\]

de Valk writes condition \ref{logweibulltail} as $\log q\in ERV_{\theta }(g)$%
.\bigskip 

\textit{Remark :} Despite its naming of log-Generalized tails, this condition also holds for Pareto
tailed distributions, as can be checked, providing $\theta =1.$

We now introduce de Valk's extreme quantile estimator. 

Let 
\[
\vartheta _{k,n}:=\sum_{j=k}^{n}\frac{1}{j}.
\]

Let $q(z)$ be the quantile of order $e^{-z}=p_{n}$ of the distribution $F$.
The estimator makes use of $X_{n-l_{n}:n}$, an intermediate order statistics
of $X_{1},..,X_{n}$, where $l_{n}$ tends to infinity \ as $n\rightarrow
\infty $ and $l_{n}$ $\ /n\rightarrow 0.$

de Valk's estimator writes 
\begin{equation}
\widehat{q}(z)=X_{n-l_{n}:n}\exp \left\{ g(\vartheta _{l_{n},n})h_{\theta
}\left( \frac{z}{\vartheta _{l_{n+1},n}}\right) \right\} .
\end{equation}

When the support of $F$ overlaps $\mathbb{R}^{-}$ then the sample size $n$
should be large; see de Valk (\cite{valk2}) for details.

Note that, in the case of a $GPD(c,a)$, parameter $\theta$ is known and equal to 1 and the normalizing function $g$ is defined by $g(x)=cx$ for $x>0$.

\subsubsection{Loss in accurracy due to binary sampling}

In Table \ref{ValkIt} we compare the performance of de Valk's method with ours on the model,
making use of complete data in de Valk's estimation, and of dichotomous ones
in our approach. Clearly de Valk's results cannot be attained by the
present sequential method, due to the loss of information induced by
thresholding and dichotomy. Despite this, the results can be compared,
since even if the bias of the estimator clearly exceeds the corresponding bias
of de Valk's, its dispersion is of the same order of magnitude, when
handling heavy tailed GPD models. Note also that given the binary nature of the data considered, the average relative error is quite honorable. 
We can assess that a large part of the volatility of the estimator produced by our sequential methodology is due to
the nature of the GPD model as well as to the sample size. 
\begin{table}[tbp]
\renewcommand{\arraystretch}{1.2} 
\begin{center}
\begin{tabular}{|c||c|c||c|c|}
\hline
& \multicolumn{4}{c|}{\textbf{Relative error on the  $(1-\alpha)-$quantile}}
\\ 
\textbf{Parameters} & \multicolumn{2}{c||}{\textbf{On complete data}} & \multicolumn{2}{c|}{\textbf{On binary data}}
\\ 
& Mean & Std & Mean & Std \\ \hline\hline
$c=0.8$, $a_0=1.5$ and $s_\alpha= 469.103$ & \hspace{0.2cm} 0.052 \hspace{0.3cm}& 0.257 & \hspace{0.2cm} -0.222 \hspace{0.2cm} & 0.554
\\ \hline
$c=1.5$, $a_0=1.5$ and $s_\alpha=31621.777 $ & 0.086 & 0.530 & -0.504 & 0.720
\\ \hline
$c=1.5$, $a_0=3$ and $s_\alpha=63243.550 $ & 0.116 & 0.625 & 0.310 & 0.590 \\ 
\hline
\end{tabular}%
\end{center}
\caption{Mean and std of the relative errors on the $1-\alpha-$quantile of GPD on complete and binary data for samples of size $n=250$ computed through $400$ replicas of both estimation procedures.\newline
Estimations on complete data are obtained with de Valk's method; estimations on binary data are provided by the sequential design.}
\label{ValkIt}
\end{table}

\section{Sequential design for the Weibull model}\label{WeibullModel}

The main property which led to the GPD model is the stability through
threshold conditioning.\ However the conditional distribution of $\widetilde{%
R}$ given $\left\{ \widetilde{R}>s\right\}$ takes a rather simple form which
allows for some variation of the sequential design method.

\subsection{The Weibull model}

Denote $\widetilde{R}\sim W(\alpha ,\beta )$, with $\alpha ,\beta >0$ a
Weibull r.v. with scale parameter $\alpha $ and shape parameter $\beta .$
let $G$ denote the distribution function of $\widetilde{R}$ , $g$ its
density function and $G^{-1}$ its quantile function. We thus write for non
negative $x$ 
\[
\begin{split}
~G(x)& =1-\exp \left( -\left( \frac{x}{\alpha }\right) ^{\beta }\right) \\
\text{ for }0<u<1,~~G^{-1}(u)& =\alpha (-\log (1-u))^{1/\beta }
\end{split}%
\]

The conditional distribution of $\widetilde{R}$ is a truncated Weibull
distribution 

\[
\begin{split}
\text{ for }\widetilde{s}_{2}>\widetilde{s}_{1},~~\mathbb{P}(\widetilde{R}>%
\widetilde{s}_{2}\mid \widetilde{R}>\widetilde{s}_{1})& =\frac{\mathbb{P}(%
\widetilde{R}>\widetilde{s}_{2})}{\mathbb{P}(\widetilde{R}>\widetilde{s}_{1})%
} \\
& =\exp \left\{ \left( -\left( \frac{s_{2}}{\alpha }\right) ^{\beta }+\left( 
\frac{s_{1}}{\alpha }\right) ^{\beta }\right) \right\} 
\end{split}%
\]

Denote $G_{s_{2}}$ the distribution function of $\widetilde{R}$ given $%
\left( \widetilde{R}>\widetilde{s}_{2}\right) $.%

\bigskip

The following result helps. For $\widetilde{s}_{2}>\widetilde{s}_{1}$,

\begin{equation}
\log \mathbb{P}(\widetilde{R}>\widetilde{s}_{2}\mid \widetilde{R}>\widetilde{%
s}_{1})=\left[ \left( \frac{\widetilde{s}_{2}}{\widetilde{s}_{1}}\right)
^{\beta }-1\right] \log \mathbb{P}(\widetilde{R}>\widetilde{s}_{1})
\end{equation}%
Assuming $\mathbb{P}(\widetilde{R}>\widetilde{s}_{1})=p$,
and given $\widetilde{s}_{1}$ we may find $\widetilde{s}_{2}$ the
conditional quantile of order $1-p$ of the distribution of $\widetilde{R}$
given $\left\{ \widetilde{R}>\widetilde{s}_{1}\right\} $. This solves the
first iteration of the sequential estimation procedure through 
\[
\log p=\left[ \left( \frac{\widetilde{s}_{2}}{\widetilde{s}_{1}}\right)
^{\beta }-1\right] \log p
\]

where the parameter $\beta $ has to be estimated on the first run of trials.

The same type of transitions holds for the iterative procedure; indeed for $%
\widetilde{s}_{j+1}>\widetilde{s}_{j}>\widetilde{s}_{j-1}$

\begin{equation}
\begin{split}
\log \mathbb{P}(\widetilde{R}>\widetilde{s}_{j+1}\mid \widetilde{R}>%
\widetilde{s}_{j})& =\left[ \frac{\log \mathbb{P}(\widetilde{R}>\widetilde{s}%
_{j+1}\mid \widetilde{R}>\widetilde{s}_{j-1})}{\log \mathbb{P}(\widetilde{R}>%
\widetilde{s}_{j}\mid \widetilde{R}>\widetilde{s}_{j-1})}-1\right] \log 
\mathbb{P}(\widetilde{R}>\widetilde{s}_{j}\mid \widetilde{R}>\widetilde{s}%
_{j-1}) \\
& =\left[ \frac{\widetilde{s}_{j-1}^{\beta }-\widetilde{s}_{j+1}^{\beta }}{%
\widetilde{s}_{j-1}^{\beta }-\widetilde{s}_{j}^{\beta }}-1\right] \log 
\mathbb{P}(\widetilde{R}>\widetilde{s}_{j}\mid \widetilde{R}>\widetilde{s}%
_{j-1})
\end{split}%
\end{equation}%
At iteration $j$ the thresholds $\widetilde{s}_{j}$ and $\widetilde{s}_{j-1}$
are known; the threshold $\widetilde{s}_{j+1}$ is the $(1-p)-$ quantile of the
conditional distribution, $\mathbb{P}(\widetilde{R}>\widetilde{s}_{j+1}\mid 
\widetilde{R}>\widetilde{s}_{j})=p$, hence solving 
\[
\log p=\left[ \frac{\widetilde{s}_{j-1}^{\beta }-\widetilde{s}_{j+1}^{\beta }%
}{\widetilde{s}_{j-1}^{\beta }-\widetilde{s}_{j}^{\beta }}-1\right] \log p
\]%
where the estimate of $\beta $ is updated from the data collected at
iteration $j.$

\subsection{Numerical results}

Similarly as in Sections \ref{Subsection numGPD} and \ref{subsection
comparaison de Valk GPD} we explore the performance of the sequential design
estimation on the Weibull model. We estimate the $(1-\alpha)-$ 
quantile of the Weibull distribution in three cases. In the first one, the
scale parameter $a$ and the shape parameter $b$ satisfy $\left( a,b\right)
=\left( 3,0.9\right)$. This corresponds to a strictly decreasing density function, with heavy tail. In the second case, the distribution is skewed since 
$\left( a,b\right) =\left( 3,1.5\right) $ and the third case is $\left( a,b\right) =\left( 2,1.5\right) $ and describes a less dispersed distribution with
lighter tail.

Table \ref{errWeibulldeValk} shows that the performance of our procedure here again depends on the shape of the distribution. The estimators are
less accurate in case 1, corresponding to a heavier tail. Those results are compared to the estimation errors on complete data through de Valk's methodology. As expected,
the loss of accuracy linked to data deterioration is similar to what was observed under the Pareto model, although a little more important. This can be explained by the fact that 
the Weibull distribution is less adapted to the splitting structure than the GPD. 

\begin{table}[tbp]
\begin{center}
\renewcommand{\arraystretch}{1.2} 
\hspace{-1cm} 
\begin{tabular}{|c||c|c||c|c|}
\hline
& \multicolumn{4}{c|}{\textbf{Relative error on the  $(1-\alpha)-$quantile}}
\\ 
\textbf{Parameters} & \multicolumn{2}{c||}{\textbf{On binary data}} & \multicolumn{2}{c|}{\textbf{On complete data}}
\\ 
& Mean & Std & Mean & Std \\ \hline\hline
$a_0=3$, $b_0=0.9$ et $s_\alpha= 25.69 $ & \hspace{0.2cm} 0.282 &\hspace{0.2cm} 0.520 & \hspace{0.2cm} 0.127 & \hspace{0.2cm}  0.197 
\\ \hline
$a_0=3$, $b_0=1.5$ et $s_\alpha=10.88 $ & -0.260 & 0.490  & 0.084 & 0.122 \\
 \hline
$a_0=2$, $b_0=1.5$ et $s_\alpha=7.25$ & -0.241 & 0.450 & 0.088 & 0.140 \\ 
\hline
\end{tabular}%
\end{center}
\caption{Mean and std of relative errors on the $(1-\alpha)-$quantile of Weibull distributions  on complete and binary data for samples of size $n=250$ computed through $400$ replicas.\newline
Estimations on complete data are obtained with de Valk's method; estimations on binary data are provided by the sequential design.}
\label{errWeibulldeValk}
\end{table}

\section{Model selection and misspecification}\label{model_selection_missp}
In the above sections, we considered two models whose presentation was mainly motivated by theoretical properties. As it has already been stated in paragraph  \ref{GPD_model}, the modeling of $\widetilde R$ by a GPD with $c$ strictly positive is justified by the assumption that the support of the original variable $R$ may be bounded by 0. However, note that the GPD model can be easily extended to the case where $c=0$. It then becomes the trivial case of the estimation of an exponential distribution. 
\smallskip

Though we did exclude the exponential case while modeling the excess probabilities of $\widetilde R$ by a GPD, we still considered the Weibull model in section \ref{WeibullModel}, which belongs to the max domain of attraction for $c=0$. On top of being exploitable in the splitting structure, the Weibull distribution is a classical tool when modeling reliability issues, it thus seemed natural to propose an adaptation of the sequential method for it. 

\smallskip
In this section, we discuss the modeling decisions and give some hints on how to deal with misspecification.

\subsection{Model selection}

The decision between the Pareto model with tail index strictly positive and the Weibull model has been covered in the literature. There exists a variety of tests on the domain of attraction of a distribution.

Dietrich and al. (2002 \cite{Dietrich2002}) Drees and al. (2006 \cite{Drees2006}) both propose a test for extreme value conditions related to Cramer-von Mises tests. Let $X$ of distribution function $G$. The null hypothesis is
\[
H_{O}: G \in MDA(c_0).
\]
In our case, the theoretical value for the tail index is $c_0=0$.
The former test provides a testing procedure based on the tail empirical quantile function, while the latter uses a weighted approximation of the tail empirical distribution. 
Choulakian and Stephens (2001 \cite{Choulakian2001}) proposes a goodness of fit test in the fashion of Cramer-von Mises tests in which the unknown parameters are replaced by maximum likelihood estimators. The test consists in two steps: firstly the estimation of the unknown parameters, and secondly the computation of the Cramer-von Mises $W^2$ or Anderson-Darling $A^2$ statistics. Let $X_1,\dots,X_n$ be a random sample of distribution $G$. The hypothesis to be tested is: $H_O$: The sample is coming from a $GPD(c_0, \widehat{a})$.  The associated test statistics are given by:
\[
\begin{split}
    &W^2 = \sum_{i=1}^n \left(\widehat{G}(x_{(i)}) - \frac{2i-1}{2n}\right)^2 + \frac{1}{12n};\\
    &A^2 = -n-\frac{1}{n} \sum_{i=1}^n (2i-1)\left\{ \log(\widehat{G}(x_{(i)})) + \log(1-\widehat{G}(x_(n+1-i)) \right\},
\end{split}
\]
where $x_{(i)}$ denotes the $i-$th order statistic of the sample. The authors provide the corresponding tables of critical points.

Jurečková and Picek (2001 \cite{Jureckova2001}) designed a non-parametric test for determining whether a distribution $G$ is light or heavy tailed. The null hypothesis is defined by :
\[
H_{c_O}:  x^{1/c_0} (1 - G(x)) \le 1 ~~  \forall x>x_0 \text{ for some } x_0>0
\]
with fixed hypothetical $c_0$.
The test procedure consists in splitting the data set in $N$ samples and computing the empirical distribution of the extrema of each sample.

\smallskip
The evaluation of the suitability of each model for fatigue data is precarious. The main difficulty here is that it is not possible to perform goodness-of-fit type tests, since firstly, we collect the data sequentially during the procedure and do not have a sample of available observations beforehand, and secondly, we do not observe the variable of interest $R$ but only peaks over chosen thresholds.
The existing tests procedures are not compatible with the reliability problem we are dealing with. On the first hand, they assume that the variable of interest is fully observed and are mainly semi-parametric or non-parametric tests based on order statistics. On the other hand, their performances rely on the availability of a large volume of data. This is not possible in the design we consider since fatigue trial are both time consuming and extremely expensive.

\smallskip
Another option consists of validating the model \textit{a posteriori}, once the procedure is completed using expert advices to confirm or not the results. For that matter, a procedure following the design presented in \ref{operational_procedure} is currently being carried out. Its results should be available in a few months and will give hints on the most relevant model.

\subsection{Misspecification}
In paragraph \ref{GPD_model}, we assumed that $\widetilde{R}$ initially follows a GPD. In practice, the distribution may have its excess probabilities converge towards it as the thresholds increase but  differ from a GPD. In the following, let us assume that $\widetilde{R}$ does not follow a GPD (of distribution function $F$) but another distribution $G$ whose tail gets closer and closer to a GPD.

\smallskip
 In this case, the issue is to control the distance between $G$ and the theoretical GPD and to determine from which thresholding level it becomes negligible. One way to deal with this problem is to restrict the model to a class of distributions that are not so distant from $F$: Assume that the distribution function $G$ of the variable of interest $\widetilde{R}$ belongs to a  neighborhood of the $GPD(c,a)$ of distribution function $F$, defined by:
\begin{equation}\label{GPD_neighborhood}
V_\epsilon(F) = \left\{G: \sup_x |\bar F(x) - \bar G(x)|w(x) \le  \epsilon\right\},
\end{equation}
where $\epsilon \ge 0$ and $w$ an increasing weight function such that $\lim_{x\rightarrow\infty} w(x) = \infty$.

$V_\epsilon(F)$ defines a neighborhood which does not tolerate large departures from $F$ in the right tail of the distribution.

\smallskip
Let $x \ge s$, it follows from \eqref{GPD_neighborhood} a bound for the conditional probability of $x$ given $R>s$:
\begin{equation}\label{cond_ineq}
\frac{\bar F(x) - \epsilon/w(x)}{\bar F(s) + \epsilon/w(s)} 
\le 
\frac{\bar G(x)}{\bar G(s)} 
\le 
\frac{\bar F(x+) + \epsilon/w(x)}{\bar F(s) - \epsilon/w(s)}.
\end{equation}
When $\epsilon=0$, the bounds of \eqref{cond_ineq} match the conditional probabilities of the theoretical Pareto distribution.

In order to control the distance between $F$ and $G$, the bound above may be rewritten in terms of relative error with respect to the Pareto distribution. Using a Taylor expansion of the right and left bounds when $\epsilon$ is close to 0, it becomes:
\begin{equation}
    1 - u(s,x).\epsilon 
     \le 
    \frac{\frac{\bar G(x)}{\bar G(s)} }{\frac{\bar F(x)}{\bar F(s)}} 
    \le 
    1 + u(x,s).\epsilon,
\end{equation}
where 
\[
u(s,x) = \frac{\left(1+\frac{cs}{a}\right)^{1/c}}{w(s)} + \frac{\left(1+\frac{cx}{a}\right)^{1/c}}{w(x)}.
\]

For a given $\epsilon$ close to 0, the relative error on the conditional probabilities can be controlled upon $s$. Indeed, then the relative error is bounded by a fixed level $\delta>0$ whenever:
\[
\frac{\left(1+\frac{cs}{a}\right)^{1/c}}{w(s)} \le \frac{\delta}{\epsilon} 
 \frac{\left(1+\frac{cx}{a}\right)^{1/c}}{w(x)}.
\]


\section{Perspectives, generalization of the two models}\label{Perspectives}
In this work, we have considered two models for $\widetilde R$ that exploits the thresholding operations used in the splitting method. This is a limit of this procedure as the lack of relevant information provided by the trials do not enable a flexible modeling of the distribution of the resistance. In the following, we present ideas of extensions and generalizations of those models, based on common properties of the GPD and Weibull models.

\subsection{Variations around mixture forms}
When the tail index is positive, the GPD is completely monotone, and thus can be written as the Laplace transform of a probability distribution.
Thyrion (1964\cite{thyrion}) and Thorin (1977\cite{thorin}) established that a $GPD(a_T,c_T)$, with $c_T>0$, can be written as the Laplace transform of a Gamma r.v $V$ whose
parameters are functions of $a_T$ and $c_T$: $V ~ \sim~ \Gamma\left(\frac{1}{c_T},\frac{a_T}{c_T}\right)$. 
Denote $v$ the density of $V$,

\begin{equation}\label{lapla}
\begin{split}
\forall x\ge 0, ~~ \bar{G}(x) = &\int_{0}^{\infty}\exp(-xy) v(y)dy \\
& \text{ where } ~~ v(y) = \frac{(a_T/c_T)^{1/c}}{\Gamma(1/c_T)}y^{1/c_T -1}\exp\left(-\frac{a_Ty}{c_T}\right).
\end{split}
\end{equation}

\medskip
It follows that the conditional survival function of  $\widetilde R$, $\bar{G}_{s_j}$, is given by:
\begin{alignat*}{2}
\mathbb{P}(\widetilde{R}>\widetilde s_{j+1}  \mid \widetilde{R}_j > \widetilde s_j) & = \bar{G}_{\widetilde s_j}(\widetilde s_{j+1}-\widetilde s_j) \\
& = \int_{0}^{\infty}\exp \left\{-(\widetilde s_{j+1} - \widetilde s_j)y \right\} v_j(y)dy,   &&\\
&\text{ ~~where }  V_j \text{ is a r.v of distribution } \Gamma\left(\frac{1}{c_j},\frac{a_j}{c_j}\right). &&
\end{alignat*}

with $c_j=c_T$ and $a_j=a_{j-1}+c_T (\widetilde s_j - \widetilde s_{j-1})$.

\medskip
Expression \eqref{lapla} gives room to an extension of the Pareto model. Indeed, we could consider distributions of $\widetilde R$ that share the same
mixture form with a mixing variable $W$ that possesses some common characteristics with the Gamma distributed r.v. $V.$

\bigskip
Similarly, the Weibull distribution $W(\alpha, \beta)$ can also be written as the Laplace transform of a stable law of density $g$ whenever 
$\beta\le1$. Indeed, it holds from Feller 1971\cite{feller}) (p. 450, Theorem 1) that:
\begin{equation}\label{fellerTh1}
\forall x\ge 0, ~~  \exp\left\{-x^{\beta} \right\}= \int_{0}^{\infty}\exp(-xy) g(y)dy 
\end{equation}
where $g$ is the density of an infinitely divisible probability distribution.
	
It follows, for $s_j< s_{j+1}$
\begin{equation}\label{condWeibullLaplace}
\begin{split}
\mathbb{P}(\widetilde{R}>\widetilde s_{j+1}  \mid \widetilde{R}_j > \widetilde s_j) &= \frac{\exp\left\{-(\widetilde s_{j+1}/\alpha)^{\beta} \right\}}{\exp\left\{-(\widetilde s_{j}/\alpha)^{\beta} \right\}} \\
& = \frac{ \int_{0}^{\infty}\exp\left\{(-(\widetilde s_{j+1}/\alpha)y\right\} g(y)dy }{ \int_{0}^{\infty}\exp\left\{-(\widetilde s_{j}/\alpha)y\right\} g(y)dy } 
= \frac{ \int_{0}^{\infty}\exp\left\{-(\widetilde s_{j+1}/\alpha)y\right\} g(y)dy }{K(s_j)} \\
&= \frac{1}{K(s_j)}  \int_{0}^{\infty}\exp\left\{-\widetilde s_{j+1}u \right\} g_{\alpha}(u)
)du \\
& \quad \quad\text{ with } u=y/\alpha \text{ and }  g_{\alpha}(u)=\alpha g(\alpha u)
\end{split}
\end{equation}

Thus an alternative modeling of $\widetilde R$ could consist in any distribution that can be written as a Laplace transform of a stable law of density $w_{\alpha,\beta}$ defined on $\mathbb{R}_+$ and parametrized by 
$(\alpha,\beta)$, that complies to the following condition:
For any $s>0$, the distribution function of the conditional distribution of $\widetilde R$ given $\widetilde R>s$ can be written as the Laplace transform of $w_{\alpha,\beta}^{(\alpha,s)}( . )$ where
\[x>s, w_{\alpha,\beta}^{(\alpha,s)}(x) = \frac{\alpha w_{\alpha,\beta}(\alpha x)}{K(s)},\] where $K( . )$ is defined in \eqref{condWeibullLaplace}.

\subsection{Variation around the GPD}
Another approach, inspired by  Naveau et al. (2016\cite{naveau2016}), consists in modifying the model so that the distribution of $\widetilde R$ tends to a GPD
as $x$ tends to infinity and it takes a more flexible form near 0.

 $\widetilde R$ is generated through $G_{(c_T,a_T)}^{-1}(U)$ with $U\sim\mathcal{U}[0,1]$. 
 Let us consider now a deformation of the uniform variable  $V=L^{-1}(U)$ defined on $[0,1]$, and the transform $W$ of the GPD: $W^{-1}(U)=G_{(c_T,a_T)}^{-1}(L^{-1} (U))$.

The survival function of the GPD being completely monotone, we can choose $W$ so that the distribution of $\widetilde R$ keeps this property.

\begin{proposition}
If $\phi : [0,\infty[ \rightarrow \mathbb{R}$ is completely monotone and let $\psi$ be a positive function, such that its derivative is completely monotone, then $\phi(\psi)$ est completely monotone.
\end{proposition}

The transformation of the GPD has cumulative distribution function $W=L(G_(c_T,a_T))$ and survival function $\bar W= \bar L(G_(c_T,a_T))$. $G(c_T,a_T)$ is a Berstein function,
thus  $\bar W$ is completely monotone if $\bar L$ is also.

\subsubsection{Examples of admissible functions:}

\emph{(1) Exponential form : }

\[
\begin{split}
&L(0) = 0 \\
& L(x) = \frac{1-\exp(-\lambda x^\alpha)}{1-\exp(-\lambda)} ~~~ \text{avec } 0\le \alpha \le 1 \text{ et } \lambda>0 \\
& L(1) = 1
\end{split}
\]

The obtained transformation is: $\forall x>0$,

\[\bar W_{(\lambda,c_T,a_T)}(x) = \bar L ( G(x)) = \frac{\exp\left(-\lambda \left[1-(1+\frac{c_T}{a_T})^{-1/c_T}\right]^\alpha \right) - \exp(-\lambda) }{1-\exp(-\lambda)} \]


with $\bar  W_{(\lambda,c_T,a_T)}(x) $ completely monotone.

\bigskip
\hspace{0.5cm} \emph{(2) Logarithmic form: }
\[
\begin{split}
&L(0) = 0 \\
& L(x) = \frac{\log(x+1)}{\log 2 } ~~~~ \text{ \big( or more generally }  \frac{\log(\alpha x+1)}{\log 2 }, ~\alpha >0 \text{\big )}\\
& L(1) = 1
\end{split}
\]

and $\forall x>0$,
\[ \bar  W_{(c_T,a_T)}(x) = 1-\frac{\log\left(2-(1+\frac{c_T}{a_T})^{-1/c_T}\right)}{\log 2}
\]

\bigskip
\hspace{0.5cm} \emph{ (3) Root form: }

\[
\begin{split}
&L(0) = 0 \\
& L(x) = \frac{\sqrt{x+1}-1 }{\sqrt{2} -1}\\
& L(1) = 1
\end{split}
\]
and 
\[ \bar  W_{(c_T,a_T)}x) = 1-\frac{\sqrt{2-(1+\frac{c_T x}{a_T})^{-1/c_T}} -1}{\sqrt{2}}
\]

\bigskip
\hspace{0.5cm} \emph{(4) Fraction form: }

\[
\begin{split}
&L(0) = 0 \\
& L(x) = \frac{(\alpha + 1)x}{x+ \alpha}, ~~ \alpha >0\\
& L(1) = 1
\end{split}
\]
and
\[ \bar W_{(\alpha,c_T,a_T)}(x) = 1-\frac{ (\alpha+1)\left(1-(1+\frac{c_T x}{a_T})^{-1/c}\right)   }{1-(1+\frac{c_T x}{a_T})^{-1/c_T} + \alpha}
\]

\begin{figure}
\begin{center}
\includegraphics[scale=0.75]{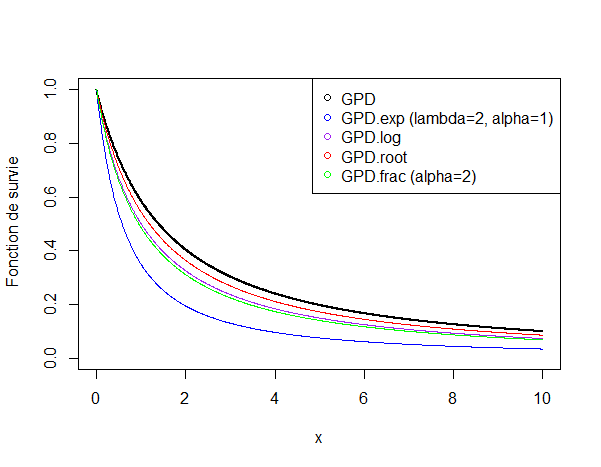}
\end{center}
\caption{Survival functions associated with transformations of the GPD$(0.8,1.5)$}
\label{transf_gpd}
\end{figure}

\bigskip
The shapes of the above transformations of the GPD are shown in Figure \ref{transf_gpd}.

\bigskip
However those transformations do not conserve the stability through thresholding of the Pareto distribution. Thus, their implementation does not give stable results. Still they give some insight on a simple generalization of the proposed models usable under additional information on the variable of interest.

\section{Conclusion}
The splitting induced procedure presented in this article proposes an innovative experimental plan to estimate an extreme quantile. Its development has been motivated by on the one hand major industrial stakes, and on the other hand the lack of relevance of existing methodologies. The main difficulty in this setting is the nature of the information at hand, since the variable of interest is latent, therefore only peaks over thresholds may be observed. Indeed, this study is directly driven from an application in material fatigue strength: when performing a fatigue trial, the strength of the specimen obviously can not be observed; only the indicator of whether or not the strength was greater than the tested level is available.

Among the methodologies dealing with such a framework, none is adapted to the estimation of extreme quantiles. We therefore proposed a plan based on splitting methods in order to decompose the initial problem into less complex ones. The splitting formula introduces a formal decomposition which has been adapted into a practical sampling strategy targeting progressively the tail of the distribution of interest. 

The structure of the splitting equation has motivated the parametric hypothesis on the distribution of the variable of interest. Two models exploiting a stability property have been presented: one assuming a Generalized Pareto Distribution and the other a Weibull distribution. 

The associated estimation procedure has been designed to use the iterative and stable structure of the model by combining a classical maximum likelihood criterion with a consistency criterion on the sequentially estimated quantiles. The quality of the estimates obtained through this procedure have been evaluated numerically. Though constrained by the quantity and quality of information, those results can still be compared to what would be obtained ideally if the variable of interest was observed. 

On a practical note, while the GPD is the most adapted to the splitting structure, the Weibull distribution has the benefit of being particularly suitable for reliability issues. The experimental campaign launched to validate the method will contribute to select a model.

\end{document}